\documentclass[11pt]{article}
\usepackage{authblk, amsmath, openbib, epsf, graphicx}
\usepackage{natbib}
\bibpunct[, ]{(}{)}{;}{a}{}{,}
\usepackage[pdftex,colorlinks=true,linkcolor=blue,citecolor=blue,urlcolor=blue]{hyperref}
\usepackage{doi}
\usepackage{amsmath,amssymb}
\usepackage{mathrsfs}
\usepackage{booktabs,dcolumn}
\usepackage{caption}
\usepackage{subcaption}
\usepackage{threeparttable}
\usepackage{booktabs}
\usepackage{rotating,booktabs}
\usepackage{ulem}  
\usepackage{indentfirst}
\usepackage{array}
\usepackage{times}
\usepackage[usenames]{color}
\usepackage{psfrag,graphicx}
\usepackage{graphicx,stackengine}
\usepackage{eso-pic,graphicx}
\usepackage{graphicx}
\usepackage{type1cm}
\usepackage{bm}
\usepackage{float}
\usepackage{multirow}
\usepackage{silence}
\usepackage{color}
\newcommand\underlay[4]{%
  \stackengine{0pt}%
  {\kern#2\includegraphics[height=#1]{#4}}%
  {\includegraphics[height=#1]{#3}}%
  {O}{l}{F}{F}{L}%
}

\newcommand{\utwi}[1]{\mbox{\boldmath $ #1$}}

\newcolumntype{z}[1]{D{.}{.}{#1}}

\newcommand{\ignore}[1]{}{}

\makeatletter    
    \if@compatibility
\renewenvironment{titlepage}
    {%
      \cleardoublepage
      \if@twocolumn
        \@restonecoltrue\onecolumn
      \else
        \@restonecolfalse\newpage
      \fi
      \thispagestyle{empty}%
    }%
    {\if@restonecol\twocolumn \else \newpage \fi
    }
\else
\renewenvironment{titlepage}
    {%
      \cleardoublepage
      \if@twocolumn
        \@restonecoltrue\onecolumn
      \else
        \@restonecolfalse\newpage
      \fi
      \thispagestyle{empty}%
    }%
    {\if@restonecol\twocolumn \else \newpage \fi
     \if@twoside\else
     \fi
    }
\fi
\makeatother

\date{}
\renewcommand{\baselinestretch}{1.5}

\usepackage[authormarkup=none, commandnameprefix=always]{changes}
\definechangesauthor[name={Taka}, color= red]{T}
\usepackage{comment}

\newcommand{\robj}[1]{{\normalfont\fontseries{b}\selectfont #1}}
\let\pkg=\textit

\begin{document}
\begin{titlepage}
\title{Tail risk forecasting with semi-parametric regression models by incorporating overnight information}

\author{
Cathy W.S. Chen\thanks{Corresponding author: Cathy W.S. Chen, E-mail: chenws@mail.fcu.edu.tw}$\:\:^{1}$,
Takaaki Koike$^{2}$, and Wei-Hsuan Shau$^{1}$
\\
$^{1}$Department of Statistics, Feng Chia University, Taiwan\\
$^{2}$Graduate School of Economics, Hitotsubashi University, Japan
}

\maketitle

\begin{abstract}
This research incorporates realized volatility and overnight information into risk models, wherein the overnight return often contributes significantly to the total return volatility.
Extending a semi-parametric regression model based on asymmetric Laplace distribution, we propose a family of RES-CAViaR-oc models by adding overnight return and realized measures as a nowcasting technique for simultaneously forecasting Value-at-Risk (VaR) and expected shortfall (ES).
We utilize Bayesian methods to estimate unknown parameters and forecast VaR and ES jointly for the proposed model family.
We also conduct extensive backtests based on joint elicitability of the pair of VaR and ES during the out-of-sample period.
Our empirical study on four international stock indices confirms that overnight return and realized volatility are vital in tail risk forecasting.

\vspace{1cm} \noindent{Keywords}: Nowcasting, Markov chain Monte Carlo method, Value-at-Risk, Expected shortfall, CAViaR model, Realized measures, Overnight return.

\vspace{1cm} \noindent{JEL Classification}: C11 $\cdot$ C22 $\cdot$ C51 $\cdot$ C53 $\cdot$ C58

\end{abstract}
\end{titlepage}

\section{Introduction}\label{sec:intro}

Value-at-Risk (VaR) is one of the most common measurements to quantify the risk of potential losses for a firm or an investment.
It presents how much money one can lose from a portfolio or stock market during a specified period.
While VaR only provides the maximum loss at a certain confidence level, it says nothing about what could happen beyond that point. In contrast, expected shortfall \citep[ES,][]{ES1999} measures the average loss in the worst-case scenarios beyond the VaR cutoff point. This feature makes ES a more comprehensive risk measure, as it better captures tail risk, which refers to the risk of extreme events.
The ES of the return $r_t$ given ${\cal{F}}_{t-1}$ at the level $\alpha$ is:
\begin{eqnarray}
ES_{t} = E(r_{t} | r_{t} < Q_{t},{\cal{F}}_{t-1}),
\end{eqnarray}
where $r_{t}$ is a daily return at time $t$, ${\cal{F}}_t$ is some information set available at time $t$, and $Q_{t}$ is the conditional VaR of $r_t$ given ${\cal{F}}_{t-1}$ at probability level $\alpha$.
ES is a coherent risk measure, whereas VaR is not due to the lack of subadditivity.
Furthermore, ES is preferred by regulators. For example, the Basel III framework recommends using both VaR and ES for market risk measurement.
In practice, both measures are often used together to get a more comprehensive understanding of risk and to develop a more robust risk management strategy. This study focuses on forecasting VaR and ES jointly during the out-of-sample period, because doing so offers a better, more efficient, and more consistent understanding of risk, assisting with decision-making processes and helping to meet regulatory demands.

With the popularity of high-frequency data, observations on intra-day returns are now more widely available.
Many studies explore intra-day return during trading hours given the availability of high-frequency data, whereby volatility can be predicted more precisely via, for example,
 realized variance \citep{RV1998,RV2003}, realized range \citep{RR12007,RR2007}, and realized kernel \citep{RK2008}.
Some contemporary works incorporate the realized volatility component into parametric models to forecast VaR and ES \citep{ASMBI, Ecosta, JFor}. \citet{GAS_RV} and \citet{RES-CAViaR2023} integrate realized volatility into the generalized autoregressive score (GAS) and the RES-CAViaR model within a semi-parametric framework. \citet{Gerlach2020} and \citet{Gerlach2023} present semi-parametric models that include realized volatility measures.

Information flow in modern financial markets is continuous, but major stock exchanges are typically only open during regular trading hours \citep{overnight2013}.
One day's opening price usually differs from the previous day's closing price, with the corresponding overnight return often accounting for a significant portion of the total daily return. This study integrates overnight information by defining the difference between today's opening price and yesterday's closing price. Consequently, we forecast VaR and ES immediately after today's market opening. In this context, the cut-off line is the market opening time, with any information beyond this time remaining unknown. This concept relates to nowcasting through overnight information.

CAViaR (Conditional Autoregressive Value at Risk) first appears in \citet{CAViaR (2004)} and is a significant development in the realm of financial econometrics as it models and predicts VaR directly.
The CAViaR model is designed to forecast VaR and does not provide an estimate for ES, which is considered a more comprehensive risk measure.
\cite{ES-CAViaR2019} proposes a joint model that estimates conditional VaR and ES simultaneously and shows its superior performance to various existing models.
We refer to this joint model as ES-CAViaR in this paper.

Motivated by the superior performance of the ES-CAViaR model in~\cite{ES-CAViaR2019}, we propose to combine it with realized volatility and the concept of nowcasting through overnight information.
We name this newly developed model as RES-VAViaR-oc.
 The contribution of this proposed model is to forecast VaR and ES by adding trading information, {including close-to-close return, realized volatility, and overnight news simultaneously}. Overnight information is critical in stock markets, mainly due to the uncertainty it introduces.
Key announcements or events often happen during non-trading hours. This could include earnings reports, geopolitical events, or policy changes. These developments can lead to a significant difference in the perceived value of a stock as well as to a gap up or down at the next market opening. Since these announcements are unpredictable, they add an element of uncertainty that we must consider when forecasting the tail risks.

From a risk management perspective, the overnight return often contributes to a significant portion of total return volatility. This additional volatility introduces further uncertainty, making it essential for us to measure and manage its exposure to overnight risks. Moreover, in a globally interconnected world nowadays, developments in foreign markets can impact domestic markets; e.g., the impacts of the COVID-19 pandemic or the conflict between Ukraine and Russia. Overnight news or economic data from other countries affect investors' sentiment and expectations, causing a price adjustment when the market opens. As different countries operate in different time zones, the unpredictability of these foreign market influences adds uncertainty.

The proposed model herein is more flexible, because it includes ES-CAViaR of \cite{ES-CAViaR2019} and RES-CAViaR of \citet{RES-CAViaR2023} as special cases.
With the relevance of the asymmetric Laplace (AL) distribution and the quantile regression, we estimate all unknown parameters and forecast tail risks jointly via the Bayesian methods. This approach has various advantages.
First, it is a more efficient and flexible way to estimate all parameters for complex models. Second, the parameter restrictions are established on the prior distribution. Third, it estimates the unknown parameters and tail risk simultaneously.

To compare the forecasting abilities among competing models, we use the violation rate (VRate) and three standard backtests. These backtests include the unconditional coverage (UC) test described by \cite{UC1995}, the conditional coverage (CC) test by \cite{CC1998}, and the Dynamic Quantile (DQ) test of \cite{CAViaR (2004)}. We use these tests to evaluate VaR performance. The VRate measures the average number of instances when the return falls below the VaR forecast, and experts widely use it to assess the accuracy of the target models. The closer VRate is to the given level $\alpha$, the better is the model's performance in forecasting VaR.
In addition to these traditional backtesting techniques, we use the quantile score~\citep{JBES2005} for the comparative backtest of VaR.

Regarding the ES evaluation, we first conduct the measurement of \cite{ES2005}, which has the benefit of directly connecting ES to VaR and to the tail of the loss distribution.
In addition, we apply the regression-based backtest proposed by~\citet{BD22} for solely backtesting ES.
For the pair assessment of VaR and ES, we consider the AL log score~\citep{ES-CAViaR2019}, which integrates the AL distribution and the class of scoring functions derived by \cite{loss function2016}.
Note that the latter two methods depend on the choice of the scoring function, for which the ranking of competing forecasts may change under a different scoring function~\citep{P20}.
Due to this concern, we provide the Murphy diagram of \citet{EGJK16} and~\citet{ZKJF20}, which is a powerful tool in assessing the quality of probabilistic forecasts of VaR or ES via visual inspection. 
The key advantages of using Murphy Diagrams for VaR or ES include
(1) robustness: provides a robust forecasting evaluation against the choice of the scoring function;
(2) comprehensive comparison: allows for performance assessment
for a relevant class of scoring functions;
(3) visual interpretation: facilitates clear graphical representation of model performance; and 
(4) versatility: applicable to various types of risk models.
We also conduct a formal hypothesis test for forecast dominance~\citep{ZKJF20}, which is a strong concept showing that one forecast is superior to another under a relevant class of scoring functions.

The \citet{BCBS2016} recommends a shift in risk metrics from VaR to ES and a reduction in the confidence level from 99\% to 97.5\%. This change heightens the focus on tail risk, thereby enabling banks to better understand their risk exposure through the examination of a more extensive set of worst-case scenarios and ensuring they have adequate capital to absorb potential losses. In alignment with these suggestions, we evaluate tail forecasts, VaR and ES, at the 1\% and 2.5\% levels for four international stock markets: NASDAQ in the U.S., DAX in Germany, HSI in Hong Kong, and Nikkei 225 in Japan.
We consider five competing risk models, three related to our proposed models with overnight information, as well as ES-CAViaR of \citet{ES-CAViaR2019} and RES-CAViaR with realized volatility of \citet{RES-CAViaR2023}.

The above-mentioned backtests and Murphy diagrams confirm that CAViaR-type models with overnight return and realized volatility are more efficient in forecasting tail risk than models without incorporating overnight information.
In particular, our Murphy diagrams and related hypothesis tests indicate a strong relation of forecast dominance~\citep{ZKJF20} between models with and without incorporating overnight information, which demonstrates the improvement of tail risk forecasting by nowcasting independently of the choice of the scoring function.
Finally, by comparing standardized score differences, we observe that the improvement of incorporating overnight information is typically more significant in RES-CAViaR compared with ES-CAViaR, and in the two Asian markets in Japan and Hong Kong compared with those in U.S. and Germany.

The rest of the paper runs as follows. Section 2 presents the RES-CAViaR-type models. Section 3 explains the Bayesian Markov Chain Monte Carlo (MCMC) method for estimating unknown parameters, and describes our forecast evaluation methods. Section 4 shows the empirical analysis, which adopts four market indices to ensure the performance of our proposed models. Section 5 concludes the study.

\section{Realized volatility CAViaR-type models}

An asset's opening price is usually not identical to its previous day's closing price, because essential information related to the listed companies might be released after the financial market closes. The difference is that after-hours trading changes investor valuations or expectations for assets.
Aside from news about companies, the development of after-hours trading has significantly influenced the difference between the previous closing price and the opening price. After-hours trading can also reflect volatility of a stock price.

When incorporating the idea of nowcasting, we take information on the difference between the opening price and the previous day's closing price into our risk model. We calculate today's tail risk once the market opens.
For a given stock of interest, $O_t$ and $C_t$ respectively denote its opening price and closing price at time $t$.
The overnight return at time $t$ is $OC_t = 100 \times \log (O_t/C_{t-1})$.
In the framework of nowcasting, the information set ${\cal{F}}^{+}_{t-1}$ is generated by the union of all closing prices up to time $t-1$ and those of opening prices until time $t$.
For a given $\alpha$ level, the conditional VaR and ES of $r_t$ given ${\cal{F}}^{+}_{t-1}$ are $Q_t$ and $ES_t$, respectively.
 Based on market reaction, we propose to use different coefficients in response to positive and negative $OC_t$.
We now describe the first proposed model as follows.

\noindent{ES-CAViaR-oc:}
\begin{eqnarray}\label{ES-CAViaR_OC}
&&Q_{t} = \beta_{1} + \beta_{2} Q_{t-1} + \beta_{3} I(OC_{t} > 0) |OC_{t}|+ \beta_{4} I(OC_{t} \leq 0)|OC_{t}|, \\
&&ES_{t} = Q_{t} - w_{t},  \nonumber\\
&&w_{t} =  \left\{\begin{array}{ll}
 \gamma_{1} + \gamma_{2} (Q_{t-1} - r_{t-1}) + \gamma_{3} w_{t-1},& \mbox{if }\: r_{t-1} \leq Q_{t-1},  \nonumber \\
 w_{t-1},& \mbox{otherwise},
\end{array}
      \right.
\end{eqnarray}
The stationarity condition in Eq.~(\ref{ES-CAViaR_OC}) is $-1<\beta_2 <1$ to ensure the stability of the time series. If the market perceives overnight information as negative, investors might become more risk-averse the following day. Additionally, if overnight information suggests that assets were previously overvalued, their prices might drop, leading to a higher VaR the next day. For these reasons, we expect $\beta_4$ to be a negative coefficient.

Realized volatility is an important factor to forecast tail risk as confirmed by many papers in the literature.
Therefore, we further include the realized volatility series $RV_{t}$ into Eq. (\ref{ES-CAViaR_OC}).

\noindent{RES-CAViaR-oc:}
\begin{eqnarray}\label{RES-CAViaR_OC}
&&Q_{t} = \beta_{1} + \beta_{2} Q_{t-1} + \beta_{3} RV_{t-1} + \beta_{4} I(OC_{t} > 0) |OC_{t}| + \beta_{5} I(OC_{t} \leq 0) |OC_{t}| , \\
&&ES_{t} = Q_{t} - w_{t},  \nonumber\\
&&w_{t} =  \left\{\begin{array}{ll}
 \gamma_{1} + \gamma_{2} (Q_{t-1} - r_{t-1}) + \gamma_{3} w_{t-1},& \mbox{if }\: r_{t-1} \leq Q_{t-1},  \nonumber \\
 w_{t-1},& \mbox{otherwise}.
\end{array}
      \right.
\end{eqnarray}
Here, $-1<\beta_2 <1$, and we expect $\beta_5<0$.

When a stock price falls in the opening market compared to the previous closing price, it creates market volatility to which traders and investors are sensitive. Hence, we only consider a negative effect of $OC_t$ in the model.

{RES-CAViaR-oc$^{-}$}:
\begin{eqnarray}\label{RES-CAViaR_OC-}
&&Q_{t} = \beta_{1} + \beta_{2} Q_{t-1} + \beta_{3} RV_{t-1} + \beta_{4} I(OC_{t} \leq 0) |OC_t|, \\
&&ES_{t} = Q_{t} - w_{t},  \nonumber\\
&&w_{t} =  \left\{\begin{array}{ll}
 \gamma_{1} + \gamma_{2} (Q_{t-1} - r_{t-1}) + \gamma_{3} w_{t-1},& \mbox{if }\: r_{t-1} \leq Q_{t-1},  \nonumber  \\
 w_{t-1},& \mbox{otherwise}.
\end{array}
      \right.
\end{eqnarray}
Here, $-1<\beta_2 <1$, and we expect $\beta_4<0$. Note that the setting $\beta_3=0$ in~\eqref{RES-CAViaR_OC} reduces to~\eqref{ES-CAViaR_OC}, and the setting $\beta_4=0$ in~\eqref{RES-CAViaR_OC} gives~\eqref{RES-CAViaR_OC-}.
Therefore, ES-CAViaR-oc in Eq.~\eqref{ES-CAViaR_OC} and RES-CAViaR-oc$^{-}$~in Eq. \eqref{RES-CAViaR_OC-} are special cases of RES-CAViaR-oc in Eq.~\eqref{RES-CAViaR_OC}.
For this reason, we describe our analysis for the most general RES-CAViaR-oc model~\eqref{RES-CAViaR_OC} in the next section.
We use the parsimonious models~\eqref{ES-CAViaR_OC} and~\eqref{RES-CAViaR_OC-} to study the effects of the dropped explanatory variables in Section~\ref{sec:empirical}.

\section{Estimation and forecast evaluations}\label{sec:estimation}

\subsection{Bayesian MCMC approach}\label{sec:bayes}

This section describes the Bayesian approach and MCMC sampling procedures
employed in estimating unknown parameters of the proposed model and for conducting the tail risk forecasting.
For brevity, we present the procedures only for the RES-CAViaR-oc model.

Let
$\utwi{\phi} = (\utwi{\beta}^{\prime}, \utwi{\gamma}^{\prime})$ in the RES-CAViaR-oc model, where $\utwi{\beta} = (\beta_{1}, \beta_{2}, \beta_{3}, \beta_{4}, \beta_{5})^{\prime}$ and $\utwi{\gamma} = (\gamma_{1}, \gamma_{2},\gamma_3)^{\prime}$.
Fundamentally, the MCMC method requires a posterior distribution of $\utwi{\phi}$, which is the product of a prior distribution $P(\utwi{\phi})$ and a likelihood function ${\cal L}(\utwi{\phi}|\mathbf{r}, \mathbf{OC}, \mathbf{RV})$, where $\textbf{r} = (r_{1},\ldots,r_{n})^{\prime}$, $\textbf{OC} = (OC_{1},\ldots,OC_{n})^{\prime}$, and $\textbf{RV} = (RV_{1},\ldots,RV_{n})^{\prime}$.
We adopt the AL distribution as the log-likelihood function:
\begin{eqnarray}\label{loglikelihoodRESCAViaROC}
\log {\cal{L}} (\mathbf{r}, \mathbf{OC}, \mathbf{RV}| \utwi{\beta}, \utwi{\gamma}) = \sum_{t=1}^{n}\left(\log \frac{\alpha - 1}{ES_{t}} + \frac{(r_{t} - Q_{t})(\alpha - I(r_{t} \leq Q_{t}))}{\alpha ES_{t}}\right).
\end{eqnarray}
As in \cite{ES-CAViaR2019}, we assume throughout the paper that $\mathbb{E}[r_t|{\cal F}_{t-1}^{+}]=0$ for every $t$.

With flat priors on $\utwi{\beta}$ and $\utwi{\gamma}$, the prior specifications go as follows.
\begin{eqnarray}
P(\utwi{\beta}) \propto I(A_1), \quad P(\utwi{\gamma}) \propto I(A_2),
\end{eqnarray}
where $A_1=\{ | \beta_{2} | < 1, \beta_3 ,\beta_5 < 0\}$, and $A_2=\{ \gamma_{1},\gamma_{2} \ge 0, \: 0 \le \gamma_{3} < 1\}$.
We impose $|\beta_2|<1$ and $0 \le \gamma_{3} < 1$ to guarantee the stability of the time series.
In addition, the downside effects of realized volatility and negative overnight return to the tail risk are incorporated into the conditions $\beta_3, \beta_5<0$.
Finally, the constraints $ \gamma_{1},\gamma_{2} \ge 0$ assure that $ES_t \le Q_t\le 0$ for every $t$.

The conditional posterior distribution is expressed by the likelihood function and prior distribution as follows:
\begin{eqnarray}
P(\utwi{\phi}_j|\utwi{r}, \utwi{OC}, \utwi{RV}, \utwi{\phi}_{-j}) \propto {\cal{L}}(\utwi{r}, \utwi{OC}, \utwi{RV} |\utwi{\phi})P(\utwi{\phi}_j|\utwi{\phi}_{-j}),
\end{eqnarray}
where $ {\cal{L}(\cdot)}$ is the likelihood function for the proposed model in the description, and $\utwi{\phi}_{-j}$ represents the vector $\utwi{\phi}$ without component $j$.

In order to estimate the {nonstandard} posterior distribution for the proposed model, we employ an adaptive MCMC algorithm of \cite{MCMC2006}, which integrates the random walk Metropolis algorithm \citep{Metropolis et al. (1953)} and independent kernel Metropolis-Hastings (MH) algorithm \citep{Hastings (1970)}. The parameter groups of $\utwi{\beta}$ and $\utwi{\gamma}$ are updated based on an adaptive MCMC method separately. The simulation study in \citet{TCAViaR2011} employs Bayesian methods for the general quantile regression problem using the asymmetric-Laplace distribution. Their approach is designed for parameter estimation of the CAViaR model family via an adaptive MCMC sampling scheme. The study demonstrates favorable estimation performance regarding precision and efficiency compared to numerical optimization of the standard quantile criterion function.
Although the model proposed in \citet{TCAViaR2011} does not factor in realized volatility and overnight information, we believe the results still attest to the effectiveness of the adaptive MCMC methods for parameter estimation.

To forecast VaR and ES in the out-of-sample period for the RES-CAViaR-oc model, we choose a one-step-ahead approach with rolling window and compute them by all unknown parameters estimated in the MCMC procedure. Let $N$ be the number of total iterations of the MCMC run and $M$ be the burn-in period.
The procedure based on the MCMC algorithm goes as follows.
\begin{description}
\item{Step 1:  } Initialize $\utwi{\phi}^{[0]} = (\utwi{\beta}^{[0]}, \utwi{\gamma}^{[0]})$.
\item{Step 2:  } For the $j$th iteration, draw from the conditional posteriors:
 \begin{eqnarray*}
 && P\left (\utwi{\beta}^{[j]}|\utwi{r}, \utwi{OC}, \utwi{RV}, \utwi{\gamma}^{[j-1]}\right )\\
 && P\left (\utwi{\gamma}^{[j]}|\utwi{r}, \utwi{OC}, \utwi{RV}, \utwi{\beta}^{[j]}\right )
 \end{eqnarray*}
 by the random walk Metropolis if $j<M$ and independent kernel MH if $j \geq M$.

\item{Step 3:  }Collect $(\utwi{\beta}^{[j]}, \utwi{\gamma}^{[j]}), Q_{n+1}^{[j]}$, and $ES_{n+1}^{[j]}$ based on:
\begin{align*}
Q_{n+1}^{[j]} &= \beta_{1}^{[j]} + \beta_{2}^{[j]} Q_{n}^{[j]} + \beta_{3}^{[j]} RV_{n} + \beta_{4}^{[j]} I(OC_{n} > 0) |OC_{n}| + \beta_{5}^{[j]} I(OC_{n} \leq 0) |OC_{n}|, \\
ES_{n+1}^{[j]} &= Q_{n+1}^{[j]} - w_{n+1}^{[j]}, \\
w_{n+1}^{[j]}  &=\begin{cases}
 \gamma_{1}^{[j]} + \gamma_{2}^{[j]} \left( Q_{n}^{[j]} - r_n \right) + \gamma_{3}^{[j]} w_{n}^{[j]},&   r_n \leq Q_{n}^{[j]},\\
 w_{n}^{[j]},& \mbox{otherwise}.
\end{cases}
\end{align*}
\item{Step 4:  } When $j=N$, we calculate:
\begin{align*}
Q_{n+1} &= \frac{1}{N-M} \sum_{j=M+1}^{N} Q_{n+1}^{[j]}\quad
\text{and}\quad
ES_{n+1} = \frac{1}{N-M} \sum_{j=M+1}^{N} ES_{n+1}^{[j]},
\end{align*}
where $Q_{n+1}^{[j]}$ and $ES_{n+1}^{[j]}$ are obtained from Step 3.
\end{description}

\subsection{Evaluation of VaR and ES forecasting} \label{sec3-2}
It is critical that financial regulators evaluate the accuracy of the proposed models in
forecasting VaR and ES since both tail risks are unobservable.
We employ various tests to evaluate the forecast performance of the proposed models, which include traditional backtests and recent comparative backtests based on loss (scoring) functions.

First, we identify the model's forecasting accuracy equal to the nominal level $\alpha$ by computing the violation rate (VRate) for quantile forecasting:
\begin{eqnarray}
\rm{VRate} = \frac{\sum_{t=n+1}^{n+m} I(r_t < Q_t)}{m},
\end{eqnarray}
where $n$ is the in-sample period, $m$ is the out-of-sample period, and $Q_t$ stands for VaR in the models.
The closer VRate is to $\alpha$, the better the performance of the model is to forecast VaR.
For a conservative evaluation of risk, we prefer VRate to be overestimated than underestimated.

Second, we employ three traditional VaR backtest procedures to properly evaluate the accuracy of the VaR forecast: UC test, CC test, and DQ test.
Both CC and DQ are joint tests where the null hypothesis consists of the independence property of the VaR violation, equivalently correct conditional violation rate for a given model, and combined with a correct UC rate.

Third, we assess whether the ES forecast is specified correctly based on various measurements.
\citet{ES2005} consider the measure $V(\alpha)   = (|V_1(\alpha)|+|V_2(\alpha)|)/2$, where $V_1(\alpha)$ is the sample mean of $\delta_t(\alpha)=r_t - ES_t(\alpha)$ over the time points in $\{n+1,\dots,n+m\}$ when (estimated) VaR violation occurs, and  $V_2(\alpha)$ is the sample mean of $\delta_t(\alpha)$ over the time points when $\delta_t(\alpha) < q(\alpha)$ with  $q(\alpha)$ being the empirical $\alpha$-quantile of $\delta_t(\alpha)$.
We prefer the smallest value of $V$ for ES in the comparisons.
We also examine the regression-based ES backtest proposed by~\citet{BD22} for ES backtesting. We carry out three versions of the ES regression (ESR) backtests: \robj{Strict ESR}, \robj{Auxiliary ESR}, and \robj{Strict Intercept}, using the \textsf{R} package \pkg{esback}~\citep{BD19}.

Under the framework of a comparative backtest~\citep{NZ17}, a possibly vector-valued risk measure forecast $\varrho_t$ is said to (empirically) dominate another $\tilde \varrho_t$ with respect to a scoring function $S$ if:
$$
\frac{1}{m}\sum_{t=n+1}^{n+m}S(\tilde \varrho_t,r_t)\le \frac{1}{m}\sum_{t=n+1}^{n+m}S(\varrho_t,r_t).
$$
We take function $S$ to be strictly consistent in the sense that the risk measure of interest is the unique minimizer of the expectation of $S$ with respect to the return.
If such a function exists, then the risk measure is called elicitable.
It is known that VaR is elicitable, and that the pair of risk measures $(\operatorname{VaR}_\alpha,\operatorname{ES}_\alpha)$, $\alpha \in (0,1)$, is (jointly) elicitable~\citep{loss function2016}.
Therefore, we evaluate the forecasting accuracy of the series $Q_t$ and the pair of series $(Q_t,ES_t)$ under certain choices of strictly consistent scoring functions.
Among others, we use the quantile score~\citep{JBES2005} for the comparative backtest of VaR:
\begin{eqnarray}\label{loss function}
S(Q_t,r_t)=
(\alpha - \mbox{I}(r_{t} \leq Q_{t})) (r_{t} - Q_{t}).
\end{eqnarray}

For the pair of VaR and ES, we consider the AL log score~\citep{ES-CAViaR2019}, which integrates the AL distribution and the class of scoring functions derived by \cite{loss function2016}.
Under the assumption that $\mathbb{E}[r_t|{\cal F}_{t-1}^{+}]=0$, the AL log score is:
\begin{eqnarray}\label{score function}
S(Q_t, ES_t,r_t) = -\log\left(\frac{\alpha - 1}{ES_{t}}\right) - \frac{(r_{t} - Q_{t})(\alpha - \mbox{I}(r_{t} \leq Q_{t}))}{\alpha ES_{t}}.
\end{eqnarray}
The AL log score is the negative logarithm of the AL distribution, and this interpretation connects the comparative backtest based on this score with the (quasi) maximum likelihood and Bayesian quantile regression frameworks.

A potential criticism of the above backtesting framework is that the scoring functions \eqref{loss function} and \eqref{score function} are just a few of innumerable strictly consistent scoring functions of $Q_t$ and $(Q_t,ES_t)$, respectively.
To address this issue, we also provide Murphy diagrams, which enable us to check whether one forecast dominates another under a relevant class of scoring functions.
\citet{EGJK16} propose the Murphy diagram for VaR, which plots the empirical elementary scores:
$$
\frac{1}{m}\sum_{t=n+1}^{n+m} (\mbox{I}(r_{t} \leq Q_{t})-\alpha) (\mbox{I}(\eta \leq Q_{t})-\mbox{I}(\eta \leq r_{t})),
$$
against $\eta\in\mathbb{R}$.
For ES, \citet{ZKJF20} propose to plot:
$$
\frac{1}{m}\sum_{t=n+1}^{n+m}
\left\{
\mbox{I}(\eta \leq ES_{t})\left(
\frac{1}{\alpha} \mbox{I}(r_{t} \leq Q_{t}) (Q_t - r_t)
-
(Q_t - \eta)
\right) + \mbox{I}(\eta \leq r_{t})(r_t-\eta)
\right\},
$$
against $\eta\in\mathbb{R}$ to evaluate the forecasting accuracy of ES.

We refer the reader to \citet{EGJK16} and  \citet{ZKJF20} for more details, such as the range of the $x$-axis of the diagram.
These Murphy diagrams provide graphical ways to check forecast dominance, where a forecast $Q_t$, or $ES_t$, dominates others, independently of the choice of scoring functions, if its curve of empirical elementary scores against $\eta$ is lower than those of others on the entire line.
The forecast dominance undergoes formal examination using the test proposed by~\citet{ZKJF20}, which is based on the stationary bootstrap~\citep{PR94}.

\section{Empirical study}\label{sec:empirical}

This study utilizes data of daily (opening and closing) prices, as well
as realized volatility data.
We collect four market indices: Nasdaq Composite (U.S.), DAX (Germany), Hang Seng Index (HSI, Hong Kong), and Nikkei 225 (Japan).
From Oxford-Man Institute of Quantitative Finance by \cite{Oxford-Man}, we download the daily returns $r_t=(\ln(C_{t}) - \ln(C_{t-1})) \times100$, where $C_t$ is defined as closing price on day $t$,  close-to-open returns $OC_t = (\ln(O_{t}) - \ln(C_{t-1})) \times 100$, where $O_t$ is defined as opening price on day $t$, and the square root of median realized volatility is $\sqrt{medRV_t}$.

\begin{figure}[ht]
\centering
\includegraphics[width=1.0\textwidth]{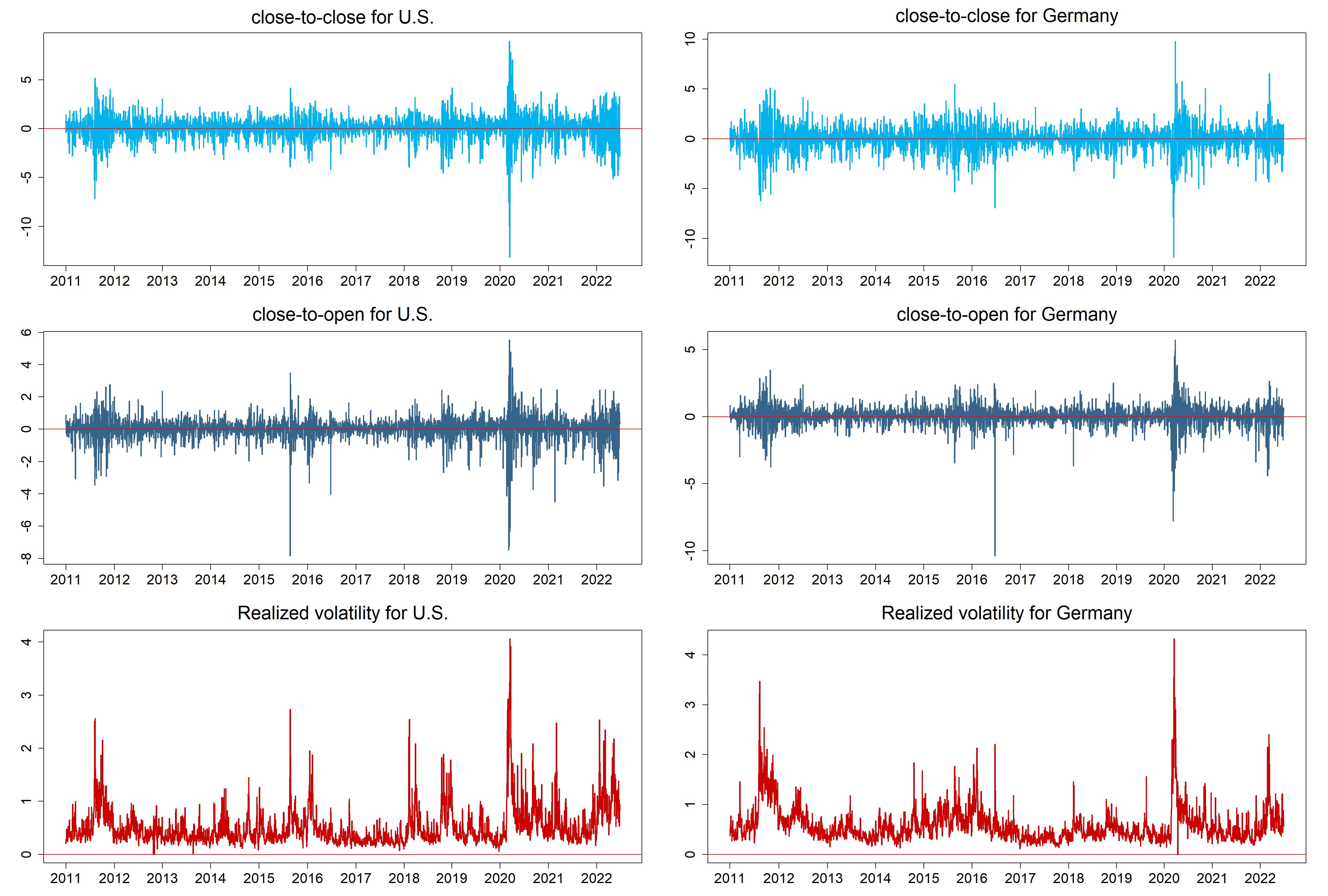}
\caption{Time plots of the U.S. and Germany stock markets}
\label{Fig1}
\end{figure}

\begin{figure}[ht]
\centering
\includegraphics[width=1.0\textwidth]{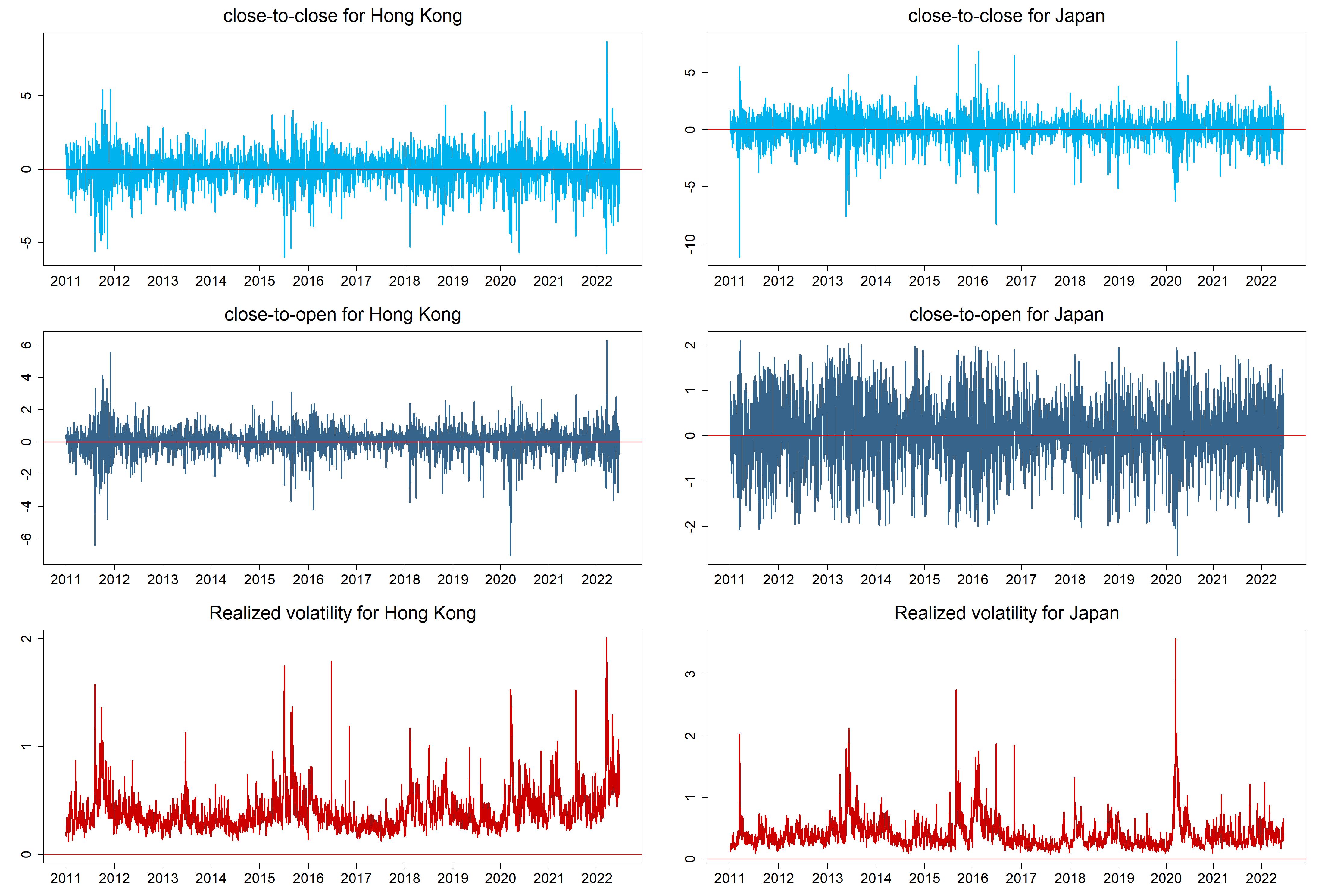}
\caption{Time plots of the Hong Kong and Japan stock markets}
\label{Fig2}
\end{figure}

We divide the dataset into two parts: one is the in-sample period from January 3, 2011 to December 31, 2017, and the other is the out-of-sample period from January 2, 2018 to June 28, 2022. The out-of-sample period covers the COVID-19 pandemic period and the circuit breakers in the U.S. stock market in March 2020.
Table \ref{summary1} exhibits the summary statistics for the series of $r_t$, $OC_t$, and $RV_t$. The series of $r_t$ and $OC_t$ exhibit left skewness, while $RV_t$ shows right skewness in all of the markets. The $OC_t$ series is much more skewed than the $r_t$ series in the out-of-sample period for all markets.
Figures \ref{Fig1} and \ref{Fig2} are the time plots of $r_t$, $OC_t$, and $RV_t$ for each market.

\begin{table}[H]
\centering
\begin{center}
\caption{\small Summary statistics of $r_t$, $OC_t$, and $RV_t$ for the four stock markets in the specified periods.}
\label{summary1}
{\scalebox{0.8}{
\begin{tabular}{llrcrrrc}
\toprule																					
Market	&	Period	&	Mean	&	Std	&	Skewness	&	Excess	    &	Min	&	Max	    \\
	    &		    &		    &		&		        &	kurtosis	&		&		 	\\																					\midrule																					
\multicolumn{1}{l}{U.S.}	&	\multicolumn{1}{l}{In-sample }	&		&		&		&		&		&		\\
	&	$r_t$	&	0.0543	&	1.0219	&	-0.4916	&	4.0615	&	-7.1685	&	5.1919		\\
	&	$OC_t$	&	0.0321	&	0.6213	&	-1.5929	&	19.3854	&	-7.8300	&	3.4956		\\
                 	&	$RV_t$	&	0.4417	&	0.2704	&	2.6949	&	12.1243	&	0.0000	&	2.7297		\\
	&	\multicolumn{1}{l}{Out-of-sample }	&		&		&		&		&		&		\\
	&	$r_t$	&	0.0430	&	1.5964	&	-0.8308	&	7.8033	&	-13.1409	&	8.9264		\\
	&	$OC_t$	&	0.0429	&	0.9676	&	-1.4884	&	11.4856	&	-7.4754	&	5.5183			\\
	&	$RV_t$	&	0.6504	&	0.4814	&	2.5585	&	10.0379	&	0.0529	&	4.0618	 	    \\
\midrule																																				
\multicolumn{1}{l}{Germany}	&	\multicolumn{1}{l}{In-sample }	&		&		&		&		&		&		\\
	&	$r_t$	&	0.0352	&	1.2463	&	-0.3188	&	2.8239	&	-6.9250	&	5.4459		\\
	&	$OC_t$	&	0.0375	&	0.6961	&	-2.0460	&	30.3964	&	-10.3707	&	3.4814		\\
                 	&	$RV_t$	&	0.6002	&	0.3483	&	2.2364	&	8.5089	&	0.1118	&	3.4666		\\
	&	\multicolumn{1}{l}{Out-of-sample }	&		&		&		&		&		&		\\
	&	$r_t$	&	0.0021	&	1.3476	&	-0.6325	&	10.2397	&	-11.8631	&	9.7634		\\
	&	$OC_t$	&	0.0218	&	0.9035	&	-0.9364	&	11.9028	&	-7.7812	&	5.7059		\\
	&	$RV_t$	&	0.5679	&	0.3766	&	4.1954	&	27.6662	&	0.0000	&	4.3226		\\
\midrule													
\multicolumn{1}{l}{Hong Kong}	&	\multicolumn{1}{l}{In-sample }	&		&		&		&		&	  &		\\
	&	$r_t$	&	0.0152	&	1.1409	&	-0.2975	&	2.7641	&	-5.9799	&	5.4535		\\
	&	$OC_t$	&	0.0553	&	0.8072	&	-0.4703	&	7.0448	&	-6.4188	&	5.5614		\\
                 	&	$RV_t$	&	0.3654	&	0.1662	&	2.7188	&	12.9573	&	0.1176	&	1.7931		\\
	&	\multicolumn{1}{l}{Out-of-sample }	&		&		&		&		&		&		\\
	&	$r_t$	&	-0.0295	&	1.3688	&	-0.0630	&	3.1543	&	-5.7351	&	8.7072		\\
	&	$OC_t$	&	0.0448	&	0.9502	&	-0.7125	&	6.4850	&	-7.0391	&	6.3041  	\\
	&	$RV_t$	&	0.4643	&	0.2088	&	2.1383	&	7.8327	&	0.1592	&	2.0100  	\\

\midrule																					
\multicolumn{1}{l}{Japan}	&	\multicolumn{1}{l}{In-sample }	&		&		&		&		&		&		\\
	&	$r_t$	&	0.0466	&	1.3680	&	-0.5936	&	6.1547	&	-11.1534	&	7.4262		\\
	&	$OC_t$	&	0.0592	&	0.8115	&	-0.1524	&	-0.1471	&	-2.0717	&	2.1097		\\
                 	&	$RV_t$	&	0.4151	&	0.2451	&	2.7385	&	12.9574	&	0.0740	&	2.7455	\\
	&	\multicolumn{1}{l}{Out-of-sample }	&		&		&		&		&		&		\\
	&	$r_t$	&	0.0160	&	1.2765	&	-0.1262	&	3.6175	&	-6.2736	&	7.7314		\\
	&	$OC_t$	&	0.0245	&	0.7656	&	-0.2236	&	0.0276	&	-2.6409	&	1.9427		\\
	&	$RV_t$	&	0.3798	&	0.2542	&	4.6227	&	38.1402	&	0.0941	&	3.5747		\\
\bottomrule 																					
\end{tabular}}}
\end{center}
\end{table}

We consider five competing risk models. Three of them are proposed in this study: (1) ES-CAViaR-oc with overnight information in Eq. (\ref{ES-CAViaR_OC});  (2) RES-CAViaR-oc with overnight information and realized volatility in Eq. (\ref{RES-CAViaR_OC}); and (3) RES-CAViaR-oc$^{-}$ with negative overnight information and realized volatility in Eq. (\ref{RES-CAViaR_OC-}). The other two we consider for comparison are: (4) ES-CAViaR of \citet{ES-CAViaR2019} and (5) RES-CAViaR with realized volatility of \cite{RES-CAViaR2023}.
 The latter two appear as follows.

\noindent\text{ES-CAViaR:}
\begin{eqnarray}\label{ES-CAViaR-Add}
&&Q_{t} = \beta_{1} + \beta_{2} I(r_{t-1} > 0) |r_{t-1}| + \beta_{3} I(r_{t-1} \leq 0) |r_{t-1}| + \beta_{4} Q_{t-1},\\
&&ES_{t} = Q_{t} - w_{t},  \nonumber\\
&&w_{t} =  \left\{\begin{array}{ll}
 \gamma_{1} + \gamma_{2} (Q_{t-1} - r_{t-1}) + \gamma_{3} w_{t-1},& \mbox{if }\: r_{t-1} \leq Q_{t-1}.\nonumber\\
 w_{t-1},& \mbox{otherwise}.
\end{array}
      \right.
\end{eqnarray}

\noindent\text{RES-CAViaR:}
\begin{eqnarray}\label{RES-CAViaR}
&&Q_{t} = \beta_{1} + \beta_{2} Q_{t-1} + \beta_{3} RV_{t-1} , \\
&&ES_{t} = Q_{t} - w_{t},  \nonumber\\
&&w_{t} =  \left\{\begin{array}{ll}
 \gamma_{1} + \gamma_{2} (Q_{t-1} - r_{t-1}) + \gamma_{3} w_{t-1},& \mbox{if }\: r_{t-1} \leq Q_{t-1},  \nonumber \\
 w_{t-1},& \mbox{otherwise}.
\end{array}
      \right.
\end{eqnarray}
Here, $|\beta_4|<1$ in~\eqref{ES-CAViaR-Add}
and $|\beta_2|<1$ in~\eqref{RES-CAViaR} are the stationarity conditions, respectively.
For all the CAViaR models, the initial values $(Q_0,ES_0)$ are set to be negative. In our experiment, the results are robust when we vary the initial values.

The adaptive MCMC method consists of two steps.  We carry out 20,000 MCMC iterations, discard the first 8,000 iterations as the burn-in period, and include only every fourth iteration in the sample period for inference. \cite{accpetance1996} demonstrate that the acceptance rate should be between 25\% to 50\% in the MCMC procedure. To ensure rapid convergence and an optimal mix of adaptive MCMC, the trace plot and autocorrelation function (ACF) plot reflect the convergence conditions. Convergence diagnostic plots are in Supplementary Materials. We discover that the ACF plots decay quickly and
that the trace plots are a good mix, denoting that the MCMC iterations reach convergence from these
plots.

\begin{table}[t]
	\centering
	\caption{\small Posterior means, medians, standard deviations, and 95\% credible intervals of the unknown parameters of the RES-CAViaR-oc model. }
	\tabcolsep=8pt \label{parameter1}
	\scalebox{0.85}
	{\begin{tabular}[l]{@{}lrrrrrrrrrr}
\toprule					
	&	\multicolumn{5}{c}{U.S.}	&	\multicolumn{5}{c}{Germany}	\\												
\cmidrule(lr){2-6} \cmidrule(lr){7-11}																		
& \multicolumn{1}{c}{Mean} & \multicolumn{1}{c}{Median} & \multicolumn{1}{c}{Std} & \multicolumn{1}{c}{2.5$\%$}   &\multicolumn{1}{c}{97.5$\%$} &\multicolumn{1}{c}{Mean} & \multicolumn{1}{c}{Median}  & \multicolumn{1}{c}{Std} & \multicolumn{1}{c}{2.5$\%$}   &\multicolumn{1}{c}{97.5$\%$} \\																					
\midrule																					
\multicolumn{1}{l}{$\beta_1$}	&	-0.7478	&	-0.7492	&	0.0335	&	-0.8113	&	-0.6799	&	-0.7199	&	-0.7161	&	0.0582	&	-0.8349	&	-0.6101	\\
\multicolumn{1}{l}{$\beta_2$}	&	0.2496	&	0.2496	&	0.0267	&	0.1981	&	0.3011	&	0.3636	&	0.3631	&	0.0262	&	0.3118	&	0.4147	\\
\multicolumn{1}{l}{$\beta_3$}	&	-1.6094	&	-1.6045	&	0.1007	&	-1.8128	&	-1.4347	&	-1.2814	&	-1.2792	&	0.0907	&	-1.4914	&	-1.1087	\\
\multicolumn{1}{l}{$\beta_4$}	&	0.0760	&	0.0779	&	0.0707	&	-0.0698	&	0.2174	&	0.4841	&	0.4866	&	0.0845	&	0.3350	&	0.6455	\\
\multicolumn{1}{l}{$\beta_5$}	&	-1.1728	&	-1.1766	&	0.0638	&	-1.2867	&	-1.0407	&	-1.0442	&	-1.0446	&	0.0263	&	-1.0982	&	-0.9910	\\
\multicolumn{1}{l}{$\gamma_1$}	&	0.2652	&	0.2658	&	0.0611	&	0.1420	&	0.3851	&	0.2807	&	0.2857	&	0.0723	&	0.1309	&	0.4154	\\
\multicolumn{1}{l}{$\gamma_2$}	&	0.2561	&	0.2528	&	0.1061	&	0.0576	&	0.4669	&	0.7706	&	0.7845	&	0.1384	&	0.4547	&	0.9860	\\
\multicolumn{1}{l}{$\gamma_3$}	&	0.1208	&	0.1009	&	0.0933	&	0.0044	&	0.3511	&	0.1384	&	0.1215	&	0.0993	&	0.0063	&	0.3648	\\
\hline

	&	\multicolumn{5}{c}{Hong Kong}	&	 \multicolumn{5}{c}{Japan}	\\
\cmidrule(lr){2-6} \cmidrule(lr){7-11}
\multicolumn{1}{l}{$\beta_1$}	&	-0.7128	&	-0.7117	&	0.0587	&	-0.8367	&	-0.6082	&	0.0356	&	0.0390	&	0.0264	&	-0.0200	&	0.0831	\\
\multicolumn{1}{l}{$\beta_2$}	&	0.2901	&	0.2931	&	0.0334	&	0.2117	&	0.3467	&	0.6361	&	0.6364	&	0.0111	&	0.6132	&	0.6572	\\
\multicolumn{1}{l}{$\beta_3$}	&	-1.5477	&	-1.5436	&	0.1223	&	-1.8302	&	-1.3341	&	-1.5112	&	-1.5192	&	0.1277	&	-1.7167	&	-1.2271	\\
\multicolumn{1}{l}{$\beta_4$}	&	0.3820	&	0.3835	&	0.0443	&	0.2912	&	0.4624	&	0.1516	&	0.1511	&	0.0233	&	0.1110	&	0.2002	\\
\multicolumn{1}{l}{$\beta_5$}	&	-1.0172	&	-1.0170	&	0.0541	&	-1.1266	&	-0.9139	&	-1.5864	&	-1.5857	&	0.0393	&	-1.6636	&	-1.5133	\\
\multicolumn{1}{l}{$\gamma_1$}	&	0.2564	&	0.2558	&	0.0836	&	0.0958	&	0.4143	&	0.9718	&	0.9713	&	0.0853	&	0.8105	&	1.1345	\\
\multicolumn{1}{l}{$\gamma_2$}	&	0.3547	&	0.3487	&	0.1760	&	0.0476	&	0.7084	&	0.3043	&	0.3004	&	0.0546	&	0.2014	&	0.4169	\\
\multicolumn{1}{l}{$\gamma_3$}	&	0.3396	&	0.3408	&	0.1228	&	0.0957	&	0.5781	&	0.0296	&	0.0210	&	0.0288	&	0.0008	&	0.1092	\\

\bottomrule																		
															
\end{tabular} }
\end{table}

For the initial value, we select $\utwi{\beta} = -0.1I_k$ and $\utwi{\gamma} = 0.1I_3$, where $k$ represents the number of parameters for
$\utwi{\beta}$ in the proposed model.
Tables \ref{parameter1} and \ref{parameter2} present the Bayesian estimates for the two oc-type models across the four stock markets.
These estimates include posterior means, medians, standard deviations, and 95\% credible intervals for the unknown parameters.
For the RES-CAViaR-oc model, we note that the estimates $\beta_4$ for the U.S. market and $\beta_1$ for the Japan market
are insignificant, as their 95\% credible intervals include zero.
Similarly, for the RES-CAViaR-oc$^-$ model,  $\beta_1$ for the Japan market is not significant.
Since all the other estimated coefficients are significant, we conclude that both realized volatility and positive overnight return explain the variation in tail risk.

\begin{table}[t]
	\centering
	\caption{\small Posterior means, medians, standard deviations, and 95\% credible intervals of the unknown parameters of the RES-CAViaR-oc$^{-}$ model. }
	\tabcolsep=8pt \label{parameter2}
	\scalebox{0.85}
	{\begin{tabular}[l]{@{}lrrrrrrrrrr}
\toprule					
	&	\multicolumn{5}{c}{U.S.}	&	\multicolumn{5}{c}{Germany}	\\										
\cmidrule(lr){2-6} \cmidrule(lr){7-11}																		
& \multicolumn{1}{c}{Mean} & \multicolumn{1}{c}{Median} & \multicolumn{1}{c}{Std} & \multicolumn{1}{c}{2.5$\%$}   &\multicolumn{1}{c}{97.5$\%$} &\multicolumn{1}{c}{Mean} & \multicolumn{1}{c}{Median}  & \multicolumn{1}{c}{Std} & \multicolumn{1}{c}{2.5$\%$}   &\multicolumn{1}{c}{97.5$\%$} \\																		
\midrule																					
\multicolumn{1}{l}{$\beta_1$}	&	-0.6282	&	-0.6298	&	0.0582	&	-0.7364	&	-0.5038	&	-0.7643	&	-0.7669	&	0.0443	&	-0.8427	&	-0.6740	\\
\multicolumn{1}{l}{$\beta_2$}	&	0.2616	&	0.2638	&	0.0316	&	0.1965	&	0.3209	&	0.1669	&	0.1642	&	0.0324	&	0.1081	&	0.2370	\\
\multicolumn{1}{l}{$\beta_3$}	&	-1.8802	&	-1.8839	&	0.1033	&	-2.0887	&	-1.6766	&	-1.9311	&	-1.9346	&	0.1240	&	-2.1701	&	-1.6422	\\
\multicolumn{1}{l}{$\beta_4$}	&	-1.2534	&	-1.2574	&	0.0743	&	-1.3953	&	-1.1151	&	-1.1340	&	-1.1336	&	0.0613	&	-1.2489	&	-1.0324	\\
\multicolumn{1}{l}{$\gamma_1$}	&	0.2978	&	0.2974	&	0.0623	&	0.1725	&	0.4248	&	0.2499	&	0.2515	&	0.0593	&	0.1297	&	0.3613	\\
\multicolumn{1}{l}{$\gamma_2$}	&	0.1703	&	0.1640	&	0.0944	&	0.0135	&	0.3751	&	0.7842	&	0.7954	&	0.1268	&	0.5125	&	0.9849	\\
\multicolumn{1}{l}{$\gamma_3$}	&	0.1211	&	0.1053	&	0.0877	&	0.0041	&	0.3254	&	0.0765	&	0.0633	&	0.0590	&	0.0030	&	0.2159	\\

\hline

	&	\multicolumn{5}{c}{Hong Kong}	&	 \multicolumn{5}{c}{Japan}	\\
\cmidrule(lr){2-6} \cmidrule(lr){7-11}																		
\multicolumn{1}{l}{$\beta_1$}	&	-0.5649	&	-0.5678	&	0.0401	&	-0.6325	&	-0.4819	&	0.0023	&	0.0045	&	0.0346	&	-0.0754	&	0.0741	\\
\multicolumn{1}{l}{$\beta_2$}	&	0.3055	&	0.3038	&	0.0248	&	0.2620	&	0.3571	&	0.7346	&	0.7354	&	0.0163	&	0.6990	&	0.7661	\\
\multicolumn{1}{l}{$\beta_3$}	&	-1.3807	&	-1.3800	&	0.0895	&	-1.5656	&	-1.2155	&	-1.0280	&	-1.0258	&	0.1253	&	-1.2884	&	-0.7893	\\
\multicolumn{1}{l}{$\beta_4$}	&	-1.2980	&	-1.2973	&	0.0387	&	-1.3835	&	-1.2223	&	-0.9925	&	-0.9954	&	0.0623	&	-1.1095	&	-0.8728	\\
\multicolumn{1}{l}{$\gamma_1$}	&	0.2369	&	0.2299	&	0.1038	&	0.0529	&	0.4489	&	0.9872	&	0.9883	&	0.0804	&	0.8327	&	1.1430	\\
\multicolumn{1}{l}{$\gamma_2$}	&	0.5217	&	0.5302	&	0.1680	&	0.1760	&	0.8435	&	0.1726	&	0.1716	&	0.0502	&	0.0732	&	0.2693	\\
\multicolumn{1}{l}{$\gamma_3$}	&	0.2908	&	0.2944	&	0.1357	&	0.0362	&	0.5493	&	0.0448	&	0.0342	&	0.0395	&	0.0013	&	0.1449	\\

\bottomrule																		
															
\end{tabular} }
\end{table}

\begin{figure}[H]
\centering
\includegraphics[width=0.9\textwidth]{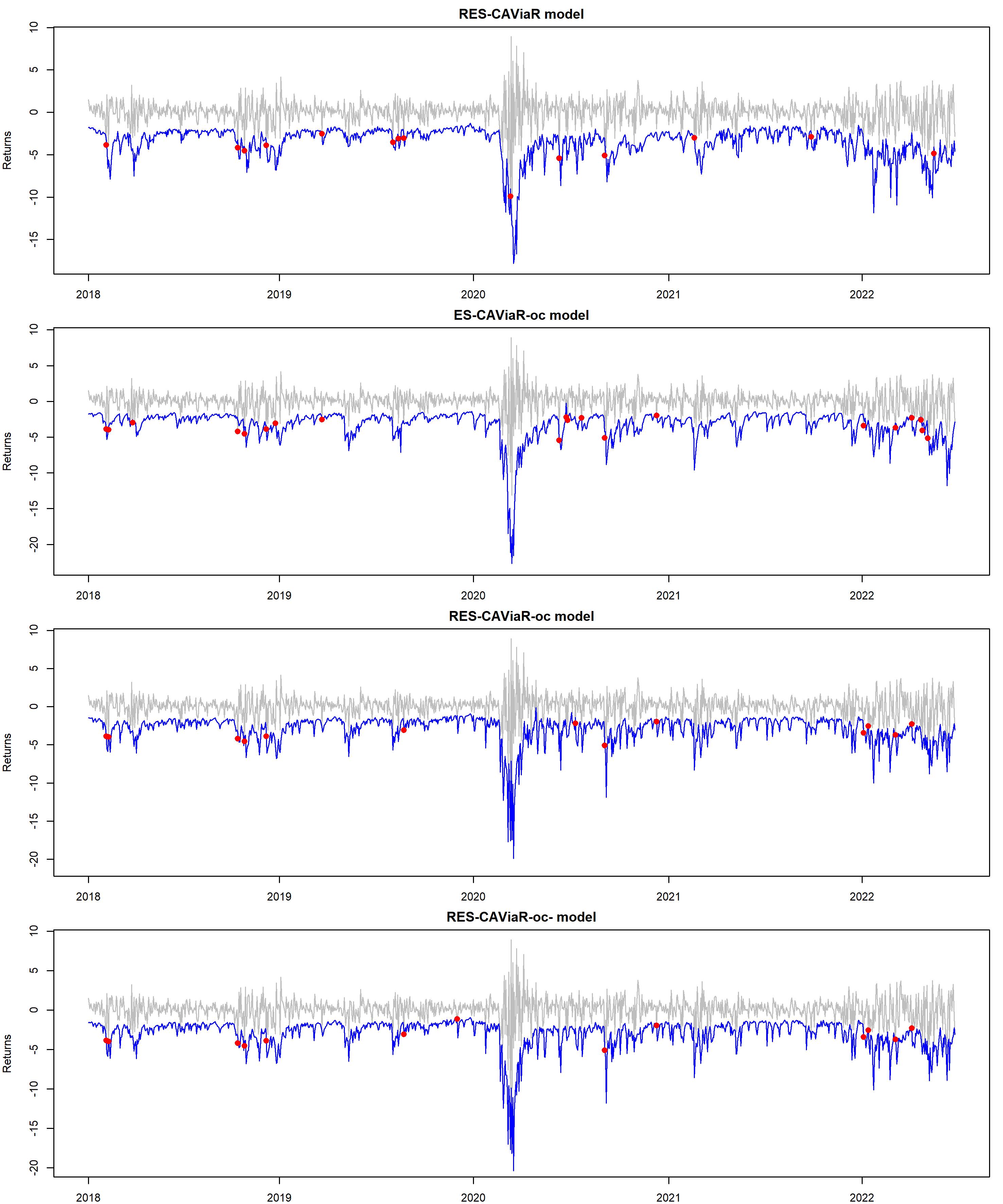}
\caption{The VaR violation plots of the U.S. market based on the RES-CAViaR, ES-CAViaR-oc, RES-CAViaR-oc, and RES-CAViaR-oc$^-$ models. Gray line represents $r_t$, blue line is 1\% VaR forecasts, and red points are the violation dates.}
\label{Fig3}
\end{figure}

\begin{figure}[H]
\centering
\includegraphics[width=0.9\textwidth]{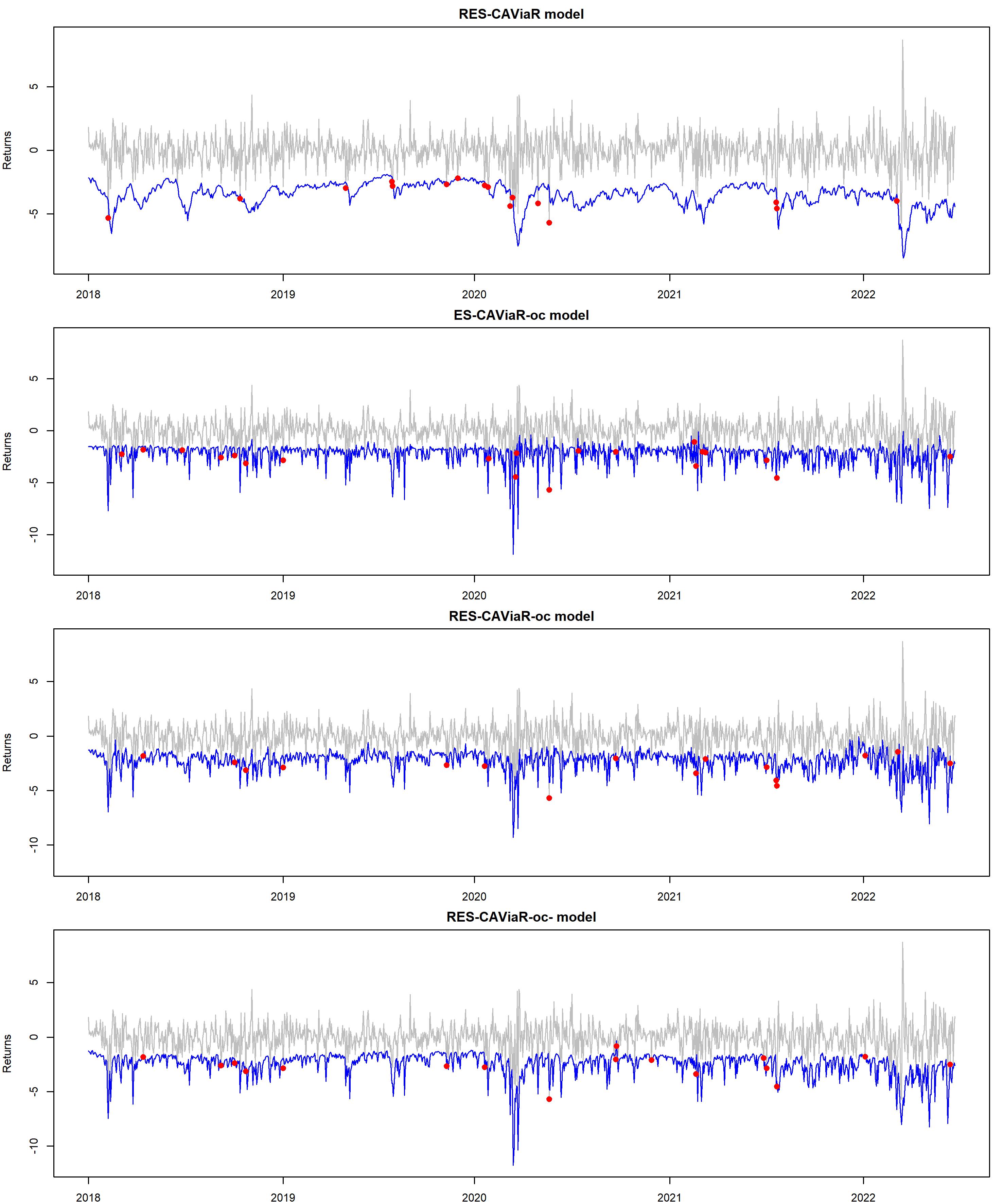}
\caption{The VaR violation plots of the Hong Kong market based on the RES-CAViaR, ES-CAViaR-oc, RES-CAViaR-oc, and RES-CAViaR-oc$^-$ models. Gray line represents $r_t$, blue line is 1\% VaR forecasts, and red points are the violation dates.}
\label{Fig4}
\end{figure}

Due to space limits, we only provide two violation plots for U.S. and Hong Kong in Figures \ref{Fig3} and \ref{Fig4}, respectively, which illustrate the VaR violation plots at the 1\% level based on the ES-CAViaR-oc, RES-CAViaR, RES-CAViaR-oc, and RES-CAViaR-oc$^-$ models.
 We examine the performance of forecasting by observing the violation during the circuit breakers in the U.S. stock market in March 2020. In this study, all four markets are successfully able to capture the extreme negative return for forecasting in the RES-CAViaR-oc and RES-CAViaR-oc$^-$ models.

\newpage

\newpage

 Table \ref{vio1} displays the 1\% level forecasting performance for each model. The fourth column shows the VRate at $\alpha=0.01$, with values closer to 1\% indicating superior models. For instance, between 0.9\% and 1.1\% VRate, 0.9\% is preferable as it forecasts more conservatively.
 The boldface values in this column highlight the models that are closest to the desired 1\% rate. The fifth column lists the rejection counts for the UC, CC, and DQ tests in out-of-sample forecasts. If the p-value is below 5\%, we count the number of rejections. Notably, both RES-CAViaR-oc and RES-CAViaR-oc$^{-}$ pass the backtests across all markets. The final column employs the ES evaluation method \citep{ES2005}, where smaller values are favored.

 In the U.S. and Germany markets, RES-CAViaR-oc$^{-}$ stands out as the best-performing model.
In the Hong Kong market, three models have the same violation rate which is the closest to 1\%. However, RES-CAViaR-oc$^{-}$ is favored when considering the ES evaluation method.
For Japan, the simple ES-CAViaR model has the violation rate closest to 1\% and the smallest ES value, making it the most favored model for this market.
Finally, the last column represents the number of rejections from three ES regression (ESR) backtesting methods \citep{BD19,BD22}: \robj{Strict ESR}, \robj{Auxiliary ESR}, and \robj{Strict Intercept}. All tests are two-sided, and decisions are based on the 10\% significance level. The results suggest that RES-CAViaR and ES-CAViaR do not forecast the $1\%$ ES effectively for the Germany market, aligning with the findings from the ES evaluation method.

Table \ref{vio2} presents the forecasting performance of each model at the 2.5\% level. According to the VRate column, at the $0.025$ level the RES-CAViaR model is most appropriate for the U.S. market, while RES-CAViaR-oc$^{-}$ is the most suitable choice for the Germany market. For Japan, RES-CAViaR-oc is selected based on its superior forecasting accuracy.
Using the ES evaluation method, the U.S. and Germany markets prefer the RES-CAViaR-oc$^{-}$ model, while the Hong Kong market opts for the ES-CAViaR-oc model. The ESR backtests reveal that the RES-CAViaR model for Hong Kong and the models incorporating overnight returns in Japan do not provide precise ES forecasts. These findings align with the results of the ES evaluation method presented in the sixth column. On the whole, the ES-CAViaR model is effective for Japan, whereas the RES-CAViaR-oc$^{-}$ model seems to be a fitting choice for the other markets.

\begin{table}[H]
\caption{\small Evaluation of 1\% VaR and ES performance during the out-of-sample periods in the four stock markets.}\label{vio1}
\begin{threeparttable}[t]
\centering
\tabcolsep=3pt
\scalebox{0.9}{
\begin{tabular}[l]{@{}clclccc}
\toprule
Market	&	Model	&	Violation 	&	Violation &	Count of rejection	&	ES & Count of rejection \\
	    &		    &	number	        &	rate \%	  &	VaR backtests$^{a}$        & evaluation$^{b}$   &	 ESR backtests$^{c}$	\\
\midrule
\multicolumn{1}{l}{U.S.}        &	{RES-CAViaR}	        &	14	&	1.25 	    &  0   &    0.1989 & 0   \\
                                &	{ES-CAViaR}	            &	22	&	1.96 	    &  3   &	0.3585	& 0\\
                                &	{ES-CAViaR-oc}	        &	20	&	1.78 	    &  3   &	0.2069  & 0 \\	
                                &	{RES-CAViaR-oc$^{-}$}	&	13	&	\bf{1.16 }	&  0   &    \bf{0.0175}  & 0 \\	
                                &	{RES-CAViaR-oc}	        &	13	&	\bf{1.16 }	&  0   &	0.0709  & 0 \\					
			
\midrule	
\multicolumn{1}{l}{Germany}     &	{RES-CAViaR}	        &	26	&	2.30 	    &  3   &	0.7876   & 1\\
                                &	{ES-CAViaR}	            &	20	&	1.77 	    &  3   &	0.6831	& 1\\
                                &	{ES-CAViaR-oc}	        &	8	&	0.71 	    &  0   &	0.1639   & 0\\
                                &	{RES-CAViaR-oc$^{-}$}	&	13	&	\bf{1.15 }	    &  0   &	\bf{0.0853} & 0  \\
                                &	{RES-CAViaR-oc}	        &	14	&	1.24 	    &  0   &	0.1374  & 0 \\							
\midrule
\multicolumn{1}{l}{Hong Kong}        &	{RES-CAViaR}	        &	16	&	1.47 	    &  2   &    0.3547	& 0\\	
                                &	{ES-CAViaR}	            &	16	&	\bf{1.47 }	    &  0   &	0.2954	& 0\\	
                                &	{ES-CAViaR-oc}	        &	20	&	1.83 	    &  3   &	0.1291	& 0\\	
                                &	{RES-CAViaR-oc$^{-}$}	&	17	&	1.56 	&  0   &	\bf{0.0825}	& 0\\	
                                &	{RES-CAViaR-oc}	        &	16	&	\bf{1.47 }	    &  0   &    0.0913	& 0\\					
\midrule		
\multicolumn{1}{l}{Japan}       &	{RES-CAViaR}	        &	15	&	1.39 	    &  0   &    0.0700	& 0\\
                                &	{ES-CAViaR}	            &	11	&	\bf{1.02 }	&  0   &	\bf{0.0137}	& 0\\		
                                &	{ES-CAViaR-oc}	        &	10	&	\bf{0.93 }	&  0   &	0.4621	& 0\\	
                                &	{RES-CAViaR-oc$^{-}$}	&	11	&	\bf{1.02 }	&  0   &	0.3034	& 0\\		
                                &	{RES-CAViaR-oc}	        &	12	&	1.11 	&  0   &	0.2094	& 0\\

\bottomrule															

\end{tabular}
}
\vspace{0.2cm}
\parbox{1.0\textwidth}{$^a$Number of rejections of UC, CC, and DQ tests are based on the 5\% significance level.\\
$^b$ The ES evaluation method by \cite{ES2005}.\\The boldface highlights the most favored model.\\
$^c$ The three ES regression (ESR) backtesting methods—\robj{Strict ESR}, \robj{Auxiliary ESR}, and \robj{Strict Intercept}, cited in \citep{BD19,BD22}—determine the number of rejections at the 10\% significance level. All tests are two-sided.}
\end{threeparttable}
\end{table}

\begin{table}[H]
\caption{\small Evaluation of 2.5\% VaR and ES performance during the out-of-sample periods in the four stock markets.}\label{vio2}
\begin{threeparttable}[t]
\centering
\tabcolsep=4pt
\scalebox{0.9}{
\begin{tabular}[l]{@{}clccccc}
\toprule
Market	&	Model	&	Violation 	&	Violation &	Count of rejection	&	ES & Count of rejection \\
	    &		    &	number	        &	rate \%	  &	VaR backtests$^{a}$        & evaluation$^{b}$   &	 ESR backtests$^{c}$	\\
\midrule															
\multicolumn{1}{l}{U.S.}		&	{RES-CAViaR}	        &	29	&	\bf{2.58\%}	&	0	&	0.2000	& 0\\
	                            &	{ES-CAViaR}	            &	43	&	3.83\%	&	2	&	0.2234	& 0\\
	                        	&	{ES-CAViaR-oc}	        &	42	&	3.74\%	&	3	&	0.1117	& 0\\
	                        	&	{RES-CAViaR-oc$^{-}$}	&	36	&	3.21\%	&	0	&	\bf{0.0505}	& 0\\
	                            &	{RES-CAViaR-oc}	        &	33	&	2.94\%	&	0	&	0.0827	& 0\\
\midrule															
\multicolumn{1}{l}{Germany}	    &	{RES-CAViaR}	        &	43	&	3.81\%	&	3	&	0.4345	& 0\\
	                            &	{ES-CAViaR}	            &	37	&	3.28\%	&	0	&	0.2255	& 0\\
	                            &	{ES-CAViaR-oc}	        &	36	&	3.19\%	&	0	&	0.3141	& 0\\
	                            &	{RES-CAViaR-oc$^{-}$}	&	30	&	\bf{2.66\%}	&	0	&	\bf{0.0394}	& 0\\
	                            &	{RES-CAViaR-oc}	        &	33	&	2.92\%	&	0	&	0.2110	& 0\\
\midrule															
\multicolumn{1}{l}{Hong Kong}	        &	{RES-CAViaR}	        &	37	&	3.39\%	&	2	&	0.2782	& 2\\
	                            &	{ES-CAViaR}	            &	41	&	3.75\%	&	2	&	0.1722	& 0\\
	                            &	{ES-CAViaR-oc}	        &	45	&	4.12\%	&	3	&	\bf{0.1350}	& 0\\
	                            &	{RES-CAViaR-oc$^{-}$}	&	30	&	\bf{2.75\%}	&	0	&	0.3566	& 0\\
	                            &	{RES-CAViaR-oc}	        &	37	&	3.39\%	&	1	&	0.2727	& 0\\
\midrule															
\multicolumn{1}{l}{Japan}	    &	{RES-CAViaR}	        &	31	&	2.88\%	&	0	&	0.0523	& 0\\
	                            &	{ES-CAViaR}	            &	23	&	2.13\%	&	0	&	\bf{0.0453}	& 0\\
	                            &	{ES-CAViaR-oc}	        &	26	&	2.41\%	&	0	&	0.3581	& 1\\
	                            &	{RES-CAViaR-oc$^{-}$}	&	30	&	2.78\%	&	0	&	0.3629	& 2\\
	                            &	{RES-CAViaR-oc}	        &	27	&	\bf{2.50\%}	&	0	&	0.1981	& 2\\
														
\bottomrule															

\end{tabular}
}
\vspace{0.2cm}
\parbox{1.0\textwidth}{$^a$Number of rejections of UC, CC, and DQ tests are based on the 5\% significance level.\\
$^b$ The ES evaluation method by \cite{ES2005}.\\The boldface highlights the most favored model.\\
$^c$ The three ES regression (ESR) backtesting methods—\robj{Strict ESR}, \robj{Auxiliary ESR}, and \robj{Strict Intercept}, cited in \citep{BD19,BD22}—determine the number of rejections at the 10\% significance level. All tests are two-sided.}
\end{threeparttable}
\end{table}

\begin{table}[H]
\caption{Quantile score for VaR at the 1\% and 2.5\% levels.}\label{Q_score}
\centering
\scalebox{0.8}{
\tabcolsep=10pt
\begin{tabular}[l]{@{}lcccccc}
\toprule									
Level	& Market & RES-CAViaR	&	ES-CAViaR	&	ES-CAViaR-oc	&	RES-CAViaR-oc$^{-}$	&	RES-CAViaR-oc	\\
\midrule																						
       &\multicolumn{1}{l}{U.S.}	     &	49.1663	    &	54.8570	    &	48.4814	&	\bf{38.6955}	&	38.8420	\\
       &\multicolumn{1}{l}{Germany}	     &	57.5574	    &	54.4840	    &	38.8935	&	\bf{37.7599}	&	37.7904	\\
1\%    &\multicolumn{1}{l}{Hong Kong}	     &	48.3234	    &	47.9249	    &	33.8620	&	32.6990	&	\bf{31.1612}	\\
       &\multicolumn{1}{l}{Japan}	     &	42.7430	    &	43.3155	    &	33.8938	&	34.3394	&	\bf{32.8257}	\\
       &\multicolumn{1}{l}{Avg loss$^{a}$}  & 49.4475	    &	50.1453	    &	38.7827	&	35.8734	&	\bf{35.1548}	\\
\midrule	
       &\multicolumn{1}{l}{U.S.}	     &	105.6847	&	111.0860	&	98.2114	&	85.0127	&	\bf{84.5575}	\\
       &\multicolumn{1}{l}{Germany}	     &	109.5971	&	104.0882	&	86.3762	&	95.8727	&	\bf{77.3612}	\\
2.5\%  &\multicolumn{1}{l}{Hong Kong}	     &	101.0733	&	103.9773	&	70.3721	&	\bf{72.1480}	&	72.3486	\\
       &\multicolumn{1}{l}{Japan}	     &	90.5812	    &	92.4619	    &	69.4997	&	\bf{63.0793}	&	63.5964	\\
       &\multicolumn{1}{l}{Avg loss$^{a}$}     & 101.7341	&	102.9034	&	81.1148	&	79.0282	&	\bf{74.4659}	\\

\bottomrule		
		
\end{tabular}}
\begin{tablenotes}
\item $^a$Avg loss is the average loss of the four stock markets in every model.
\item $^{*}$Boldface number represents the best model in each market.
\end{tablenotes}

\end{table}

Table \ref{Q_score} illustrates the quantile score for VaR at two different levels: 1\% and 2.5\%, for five models, in which the most accurate model should minimize the scoring functions.
For each of the two VaR levels, the lowest quantile score (boldface number) in each market indicates the best-performing model for that market, as a lower score indicates a better fit to the data.
At the 1\% VaR level, the RES-CAViaR-oc$^-$ model performs best in the U.S. and Germany markets, while the RES-CAViaR-oc model outperforms in the Hong Kong and Japan markets.
At the 2.5\% VaR level, the RES-CAViaR-oc model provides the best performance in the U.S. and Germany markets, whereas the RES-CAViaR-oc$^-$ model is superior in the Hong Kong and Japan markets.

The AL log score, as per \citet{ES-CAViaR2019}, is a measure used to evaluate the goodness of fit of these models. Lower AL log scores indicate a better fit of the model to the data.
Table \ref{AL_log_score} presents the scoring function by the AL distribution at the 1\% and 2.5\% levels, evaluating VaR and ES jointly. At the 1\% level, as the scoring function is smallest, RES-CAViaR-oc$^-$ is outstanding in the U.S. and Germany markets; otherwise, RES-CAViaR-oc is the best in the Hong Kong and Japan markets. The last row demonstrates RES-CAViaR-oc is most appropriate by average loss. As for the 2.5\% level, RES-CAViaR-oc$^-$ has the best performance in the U.S., Hong Kong, and Japan markets, and RES-CAViaR-oc has the best performance in the Germany market. Finally, the RES-CAViaR-oc is more outstanding for both functions than the others.

\begin{table}[H]
\caption{ AL log score for VaR and ES at the 1\% and 2.5\% levels.}\label{AL_log_score}
\scalebox{0.8}{
\tabcolsep=10pt
\begin{tabular}{lcccccc}
\toprule									
Level	&	Market & RES-CAViaR	&	ES-CAViaR	&	ES-CAViaR-oc	&	RES-CAViaR-oc$^{-}$	&	RES-CAViaR-oc	\\
\midrule											
        &\multicolumn{1}{l}{U.S.}	    &	2685.305	&	2984.366	&	2822.656	&	\bf{2423.667}	&	2457.262	\\
        &\multicolumn{1}{l}{Germany}	&	2942.582	&	2877.914	&	2524.169	&	\bf{2487.010}	&	2500.123	\\
1\%     &\multicolumn{1}{l}{Hong Kong}	    &	2740.954	&	2735.130	&	2378.305	&	2305.682	&	\bf{2248.653}	\\
        &\multicolumn{1}{l}{Japan}	    &	2532.623	&	2565.484	&	2325.988	&	2360.695	&	\bf{2240.589}	\\
        &\multicolumn{1}{l}{Avg loss$^{a}$}   &	2725.366	&	2790.724	&	2512.780	&	2394.264	&	\bf{2361.657}	\\
\hline
        &\multicolumn{1}{l}{U.S.}	    &	2511.184	&	2690.685	&	2492.153	&	\bf{2301.253}	&	2318.049	\\
        &\multicolumn{1}{l}{Germany}	&	2613.412	&	2568.101	&	2885.410	&	2855.559	&	\bf{2320.561}	\\
2.5\%   &\multicolumn{1}{l}{Hong Kong}	    &	2529.315	&	2565.263	&	2190.503	&	\bf{2166.356}	&	2168.694	\\
        &\multicolumn{1}{l}{Japan}	    &	2374.241	&	2399.931	&	2112.757	&	\bf{1998.683}	&	2009.564	\\
        &\multicolumn{1}{l}{Avg loss$^{a}$}   &	2507.038	&	2555.995	&	2420.206	&	2330.463	&	\bf{2204.217}	\\

\bottomrule									
\end{tabular}}
\begin{tablenotes}
\item $^a$Avg loss is the average loss of the four stock markets in each model.
\item $^{*}$Boldface number represents the best model in each market.
\end{tablenotes}

\end{table}

\begin{figure}[H]
\centering
\scalebox{0.8}{
\includegraphics[width=0.8\textwidth]{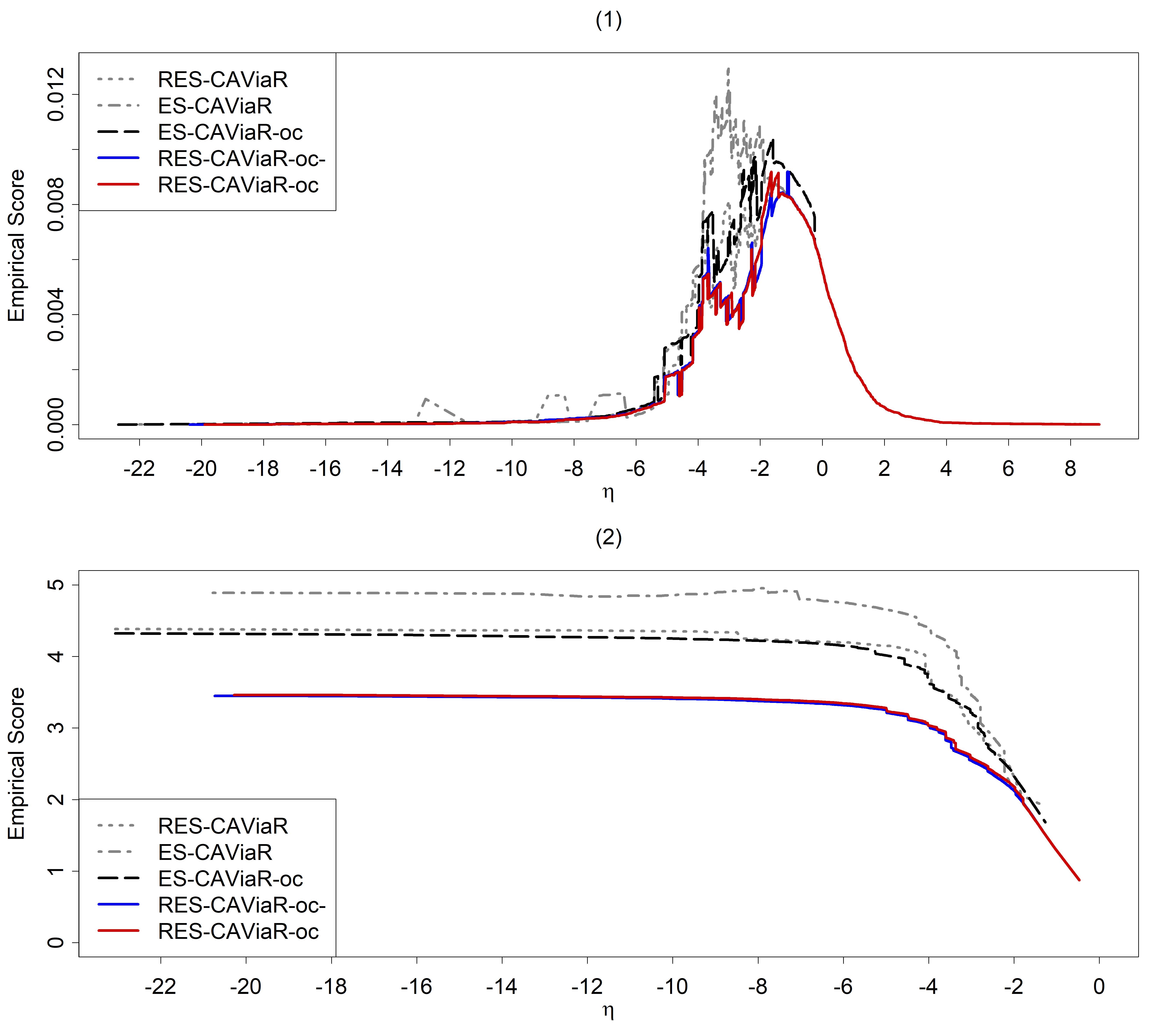}
}
\caption{Murphy diagrams for (1) VaR and (2) ES at the $1\%$ level for the U.S. market.}
\label{Fig5}
\end{figure}

\begin{figure}[H]
\centering
\scalebox{0.8}{
\includegraphics[width=0.8\textwidth]{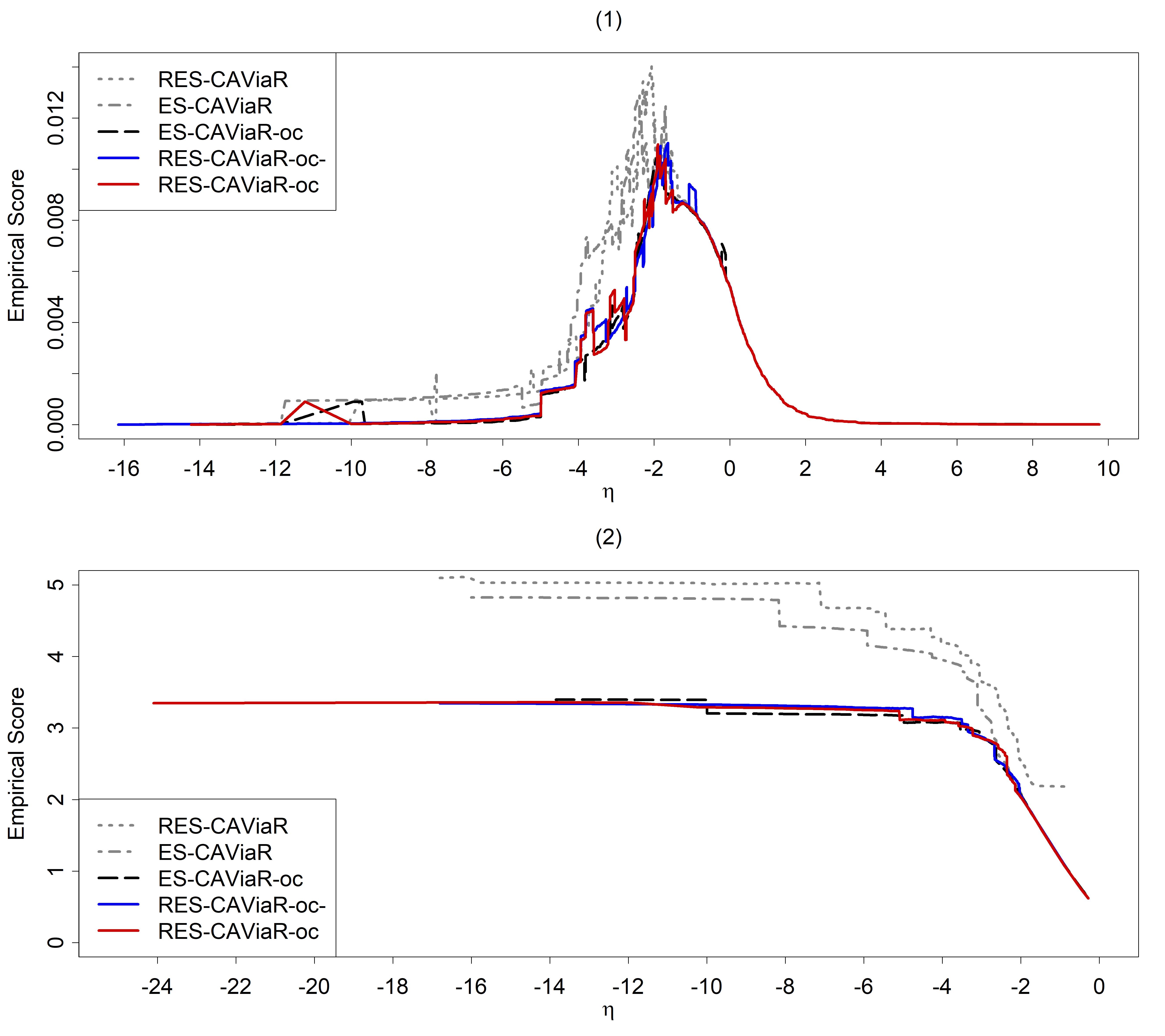}
}
\caption{Murphy diagrams for (1) VaR and (2) ES at the $1\%$ level for the Germany market.}
\label{Fig6}
\end{figure}

\begin{figure}[H]
\centering
\scalebox{0.8}{
\includegraphics[width=0.8\textwidth]{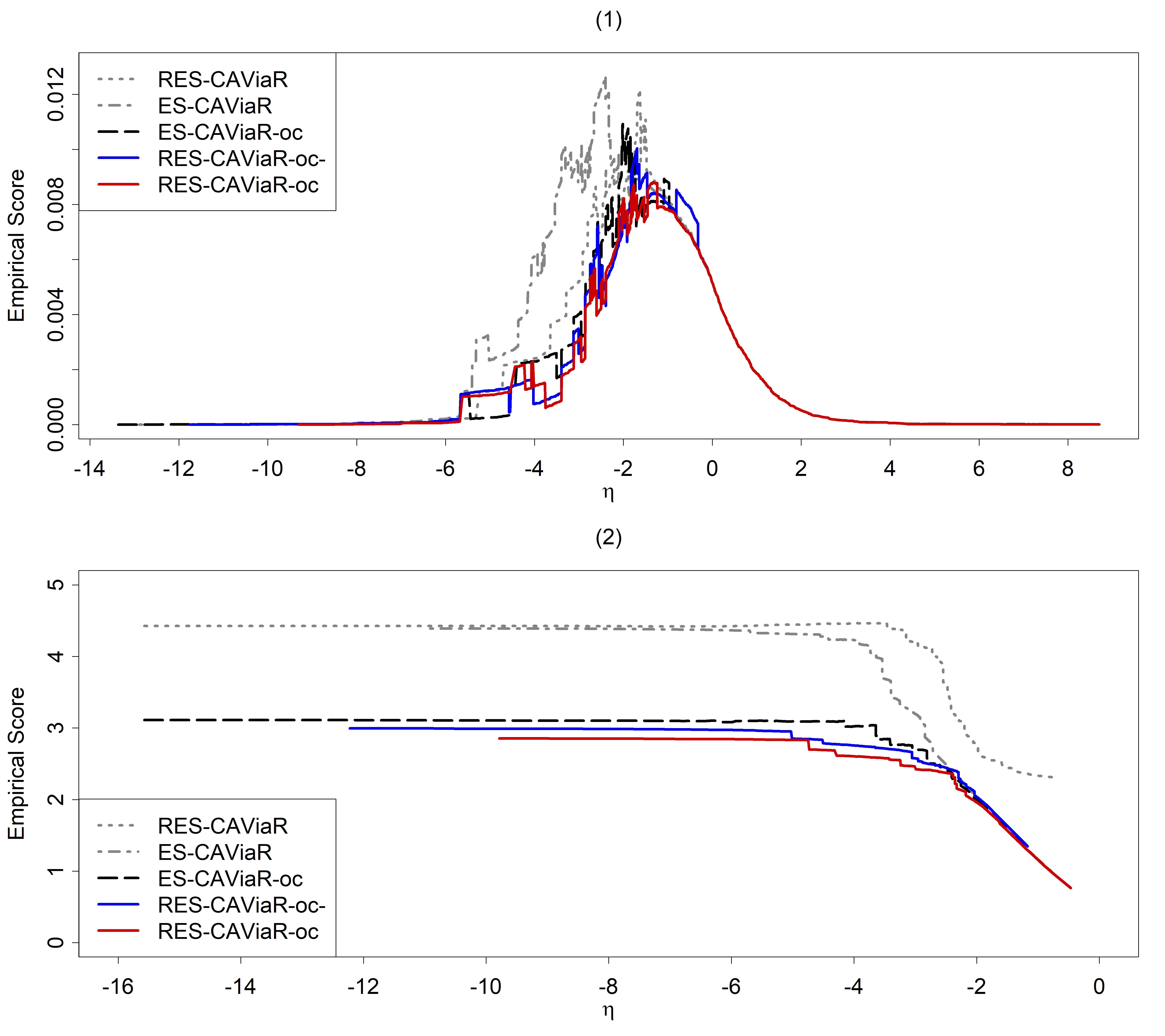}
}\caption{Murphy diagrams for (1) VaR and (2) ES at the $1\%$ level for the Hong Kong market.}
\label{Fig7}
\end{figure}

\begin{figure}[H]
\centering
\scalebox{0.8}{
\includegraphics[width=0.8\textwidth]{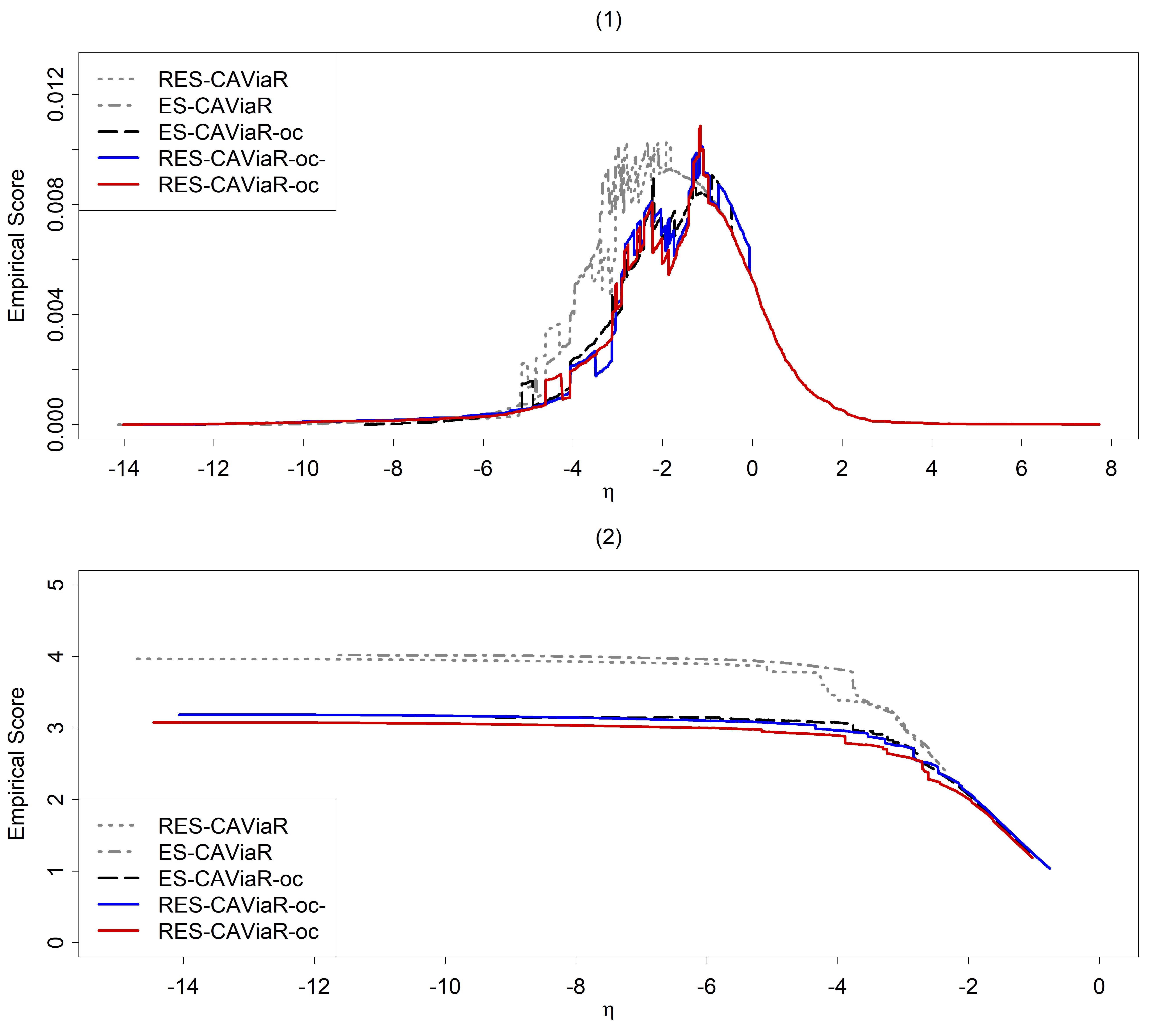}
}\caption{Murphy diagrams for (1) VaR and (2) ES at the $1\%$ level for the Japan market.}
\label{Fig8}
\end{figure}

Figures~\ref{Fig5}--\ref{Fig8} display the Murphy diagrams for the 1\% VaR and ES. From these figures, it is evident that the proposed RES-CAViaR-oc type models outperform other models, regardless of the scoring functions applied. We choose to include the Murphy diagrams for the 2.5\% VaR and ES in the Supplementary Materials, as they exhibit the same patterns as those mentioned above.
We also formally test the forecast dominance of the proposed RES-CAViaR-oc model over other models through the test proposed by~\citet{ZKJF20}.
Specifically, for each competing forecast of VaR or ES in comparison with RES-CAViaR-oc, we establish a null hypothesis: RES-CAViaR-oc outperforms the other model across the set of elementary scoring functions. If this test is rejected, then the proposed RES-CAViaR-oc model provides less accurate forecasts than its competitor, when accuracy is gauged using an elementary scoring function. Notably, for each competing forecast in comparison to RES-CAViaR-oc and for each market, the p-value is so close to $1$ that we opt not to report the findings. As a result, the hypothesis remains unchallenged even at a significance level of $10\%$.

These results formally corroborate the insights gained from the Murphy diagrams, suggesting that the proposed model outperforms others, irrespective of the scoring functions chosen. Given the extent of this dominance, as described in~\citet{EGJK16} and~\citet{NZ17}, we believe these findings strongly support the choice of the AL log score and highlight the benefits of nowcasting based on overnight information.


Tables \ref{rank1} and \ref{rank2} present summaries of model comparisons, which are ranked by five criteria at the 1\% and 2.5\% levels, respectively. The top-performing model is assigned a rank of 1, and the ranking continues in ascending order. In case of a tie, the models share the same rank, and the next rank is skipped. The last row in each market's table represents the sum of the previous four rank rows, which are ordered according to VRate, the ES method, quantile score, AL log score, and ESR backtest.

At the 1\% level, Table \ref{rank1} showcases the performance of five different models based on five criteria across four stock markets. Across the board, RES-CAViaR-oc$^-$ appears to be the most consistently high-performing model, particularly in the U.S. and Germany markets.
The RES-CAViaR-oc model exhibits solid performance across multiple markets, particularly in the U.S., Germany, and Japan markets (Table \ref{rank2}).
In reference to the last row in Tables \ref{rank1} and \ref{rank2}, the RES-CAViaR-oc$^{-}$ and RES-CAViaR-oc models are the most appropriate, as indicated by the smallest rank sum at the 1\% and 2.5\% levels, respectively. In conclusion, the ES-CAViaR model, which incorporates realized volatility and overnight return, along with RES-CAViaR-oc$^{-}$ and RES-CAViaR-oc, demonstrates strong performance in the model comparison process.

We observe that nowcasting enhances forecasting accuracy. Our next inquiry is to determine if and how this improvement varies across different models and markets. To investigate this, we examine the sample mean, denoted by $\mu$, of the score difference
$\delta_t = S_{t,2} - S_{t,1}$ over the out-of-sample period $t=n+1,\dots,n+m$, where $S_{t,1}$ presents the score of a forecast from the ES-CAViaR-oc (or RES-CAViaR-oc) model at time $t$ and $S_{t,2}$ is that of the ES-CAViaR (or RES-CAViaR) model, which does not incorporate overnight information.
We choose the quantile score for VaR and AL log score for ES.
To compare this average difference across countries, we compute the $t$-statistics $\sqrt{m}\mu / \sigma$ where $\sigma$ is an autocorrelation-consistent estimator of the standard deviation as computed in~\citet{ZKJF20}.
The results appear in Tables~\ref{imp1} and~\ref{imp2}.
The larger the reported standardized score difference is, the more nowcasting enhances forecasting accuracy.
We note that both models, especially RES-CAViaR, benefit from incorporating overnight information. Furthermore, the two Asian markets, Japan and Hong Kong, show more significant nowcasting improvement than do the U.S. and Germany.
This implication suggests that Asian markets might be more influenced by other markets that are trading live while they are closed.

\begin{table}[H]
\caption{\small Ranking of the five models based on five criteria at the 1\% level. The highest-performing model receives the lowest rank.}
\label{rank1}
\scalebox{0.8}{
\tabcolsep=7pt
\begin{tabular}{clccccc}
\toprule
Market	&	Rule	&	{RES-CAViaR}	&	{ES-CAViaR}	&	{ES-CAViaR-oc}	&	{RES-CAViaR-oc$^-$}	&	{RES-CAViaR-oc}	\\	
\midrule															
\multicolumn{1}{l}{U.S.}		&	VRate	                &	3	&	5 	&	4	&	1	&	1	\\		
	                            &	ES method               &	3	&	5 	&	4	&	1	&	2	\\		
	                            &	Quantile score	        &	4	&	5 	&	3	&	1	&	2	\\		
	                            &	AL log score            &	3	&	5 	&	4	&	1	&	2	\\
                                &   ESR backtest            &   1   &   1   &   1   &   1   &   1   \\
\cmidrule(lr){2-7}		
	                            &	Sum$^{a}$	            &	14	&	21	&	16	&	\bf{5}	&	8	\\		
\midrule															
\multicolumn{1}{l}{Germany}	    &	VRate	                &	5	&	4 	&	3	&	1	&	2	\\		
	                            &	ES method               &	5	&	4 	&	3	&	1	&	2	\\		
	                            &	Quantile score	        &	5	&	4 	&	3	&	1	&	2	\\		
	                            &	AL log score            &	5	&	4 	&	3	&	1	&	2	\\
                                &   ESR backtest             &   4  &   4   &   1   &   1   &   1   \\
\cmidrule(lr){2-7}			
	                            &	Sum	                    &	24	&	20	&	13	&	\bf{5}	&	9	\\		
\midrule															
\multicolumn{1}{l}{Hong Kong}	        &	VRate	                &	4	&	1 	&	5	&	3	&	1	\\		
	                            &	ES method               &	5	&	4 	&	3	&	1	&	2	\\		
	                            &	Quantile score	        &	5	&	4 	&	3	&	2	&	1	\\		
	                            &	AL log score            &	5	&	4 	&	3	&	2	&	1	\\
                                &   ESR backtest              &   1   &   1   &   1   &   1   &   1   \\
\cmidrule(lr){2-7}			
	                            &	Sum	                    &	20	&	14	&	15	&	9	&	\bf{6}	\\		
\midrule															
\multicolumn{1}{l}{Japan}	    &	VRate	                &	5	&	1 	&	3	&	1	&	4	\\		
	                            &	ES method               &	2	&	1 	&	5	&	4	&	3	\\		
	                            &	Quantile score	        &	4	&	5 	&	2	&	3	&	1	\\		
	                            &	AL log score            &	4	&	5 	&	2	&	3	&	1	\\
                                &  ESR backtest              &   1   &   1   &   1   &   1   &   1   \\
\cmidrule(lr){2-7}			
	                            &	Sum	                    &	16	&	13	&	13	&	12	&	\bf{10}	\\
\midrule		
\multicolumn{1}{l}{Total$^{b}$}	&		                    &	74	&	68	&	57	&	\bf{31}	&	33	\\																
\bottomrule															

\end{tabular}}
\vspace{0cm}
\begin{tablenotes}
\item {\footnotesize{$^a$Sum is the summation of five criteria.}}
\item {\footnotesize{$^b$Total is the total ranking of the four stock markets in each model.}}
\item {\footnotesize{$^{*}$Boldface number represents the best model in each market.}}
\end{tablenotes}
\end{table}

\begin{table}[H]
\caption{\small Ranking of the five models based on five criteria at the 2.5\% level. The highest-performing model receives the lowest rank.}\label{rank2}
\scalebox{0.8}{
\tabcolsep=7pt
\begin{tabular}{clccccc}
\toprule
Market	&	Rule	&	{RES-CAViaR}	&	{ES-CAViaR}	&	{ES-CAViaR-oc}	&	{RES-CAViaR-oc$^-$}	&	{RES-CAViaR-oc}	\\	
\midrule															
\multicolumn{1}{l}{U.S.}		&	VRate	                    &	1	&	4 	&	5	&	3	&	2	\\		
	                            &	ES method                   &	4	&	5 	&	3	&	1	&	2	\\		
	                        	&	Quantile score	            &	4	&	5 	&	3	&	2	&	1	\\		
	                        	&	AL log score 	            &	4	&	5 	&	3	&	1	&	2	\\
                                &   ESR backtest                &   1   &   1   &   1   &   1   &   1   \\
\cmidrule(lr){2-7}		
	                            &	Sum$^{a}$	                &	14	&	20	&	15	&	\bf{8}	&	\bf{8}	\\		
\midrule															
\multicolumn{1}{l}{Germany}	    &	VRate	                    &	5	&	4 	&	3	&	1	&	2	\\		
	                            &	ES method	                &	5	&	3 	&	4	&	1	&	2	\\		
	                            &	Quantile score	            &	5	&	4 	&	2	&	3	&	1	\\		
	                            &	AL log score 	            &	3	&	2 	&	5	&	4	&	1	\\
                                &   ESR backtest                 &   1   &   1   &   1   &   1   &   1   \\
\cmidrule(lr){2-7}		
	                            &	Sum	                        &	19	&	14	&	15	&	10	&	\bf{7}	\\		
\midrule															
\multicolumn{1}{l}{Hong Kong}	        &	VRate	                    &	3	&	4 	&	5	&	1	&	2	\\		
	                            &	ES  method                  &	4	&	2 	&	1	&	5	&	3	\\		
	                            &	Quantile score              &	4	&	5 	&	1	&	2	&	3	\\		
	                            &	AL log score                &	4	&	5 	&	3	&	1	&	2	\\
                                &   ESR backtest                 &   5   &   1   &   1   &   1   &   1   \\
\cmidrule(lr){2-7}		
	                            &	Sum	                        &	20	&	17	&	11	&	\bf{10}	&	11	\\		
\midrule															
\multicolumn{1}{l}{Japan}	    &	VRate	                    &	5	&	4 	&	2	&	3	&	1	\\		
	                            &	ES method                   &	2	&	1 	&	4	&	5	&	3	\\		
	                            &	Quantile score	            &	4	&	5 	&	3	&	1	&	2	\\		
	                            &	AL log score                &	4	&	5 	&	3	&	1	&	2	\\
                                &   ESR backtest                 &   1   &   1   &   3   &  4  &  4   \\
\cmidrule(lr){2-7}		
	                            &	Sum	                        &	16	&	16	&	15	&	14	&	\bf{12}	\\	
\midrule	
\multicolumn{1}{l}{Total$^{b}$}	&		                        &	69	&	67	&	56	&	42	&	\bf{38}	\\

\bottomrule															

\end{tabular}}
\vspace{0cm}
\begin{tablenotes}
\item {\footnotesize{$^a$Sum is the summation of five criteria.}}
\item {\footnotesize{$^b$Total is the total ranking of four stock markets in each model.}}
\item {\footnotesize{$^{*}$Boldface number represents the best model in each market.}}
\end{tablenotes}
\end{table}

\begin{table}[H]
\caption{\small Improvement of nowcasting of 1\% VaR and ES performance measured by the standardized score difference.}\label{imp1}
\centering
\tabcolsep=10pt
\begin{tabular}[l]{@{}l | cc|cc}
\toprule
&   \multicolumn{2}{c|}{VaR} &    \multicolumn{2}{c}{ES} \\ \hline
Market & ES-CaViaR & RES-CaViaR& ES-CaViaR & RES-CaViaR\\ \hline
U.S. & 1.34 & 3.87 & 0.75 & 3.19 \\ 
  Germany & 2.85 & 2.55 & 2.50 & 2.30 \\ 
  Hong Kong & 2.83 & 5.14 & 2.30 & 4.59 \\ 
  Japan & 3.78 & 4.13 & 2.85 & 4.06 \\ 
\bottomrule															
\end{tabular}
\end{table}

\begin{table}[H]
\caption{\small Improvement of nowcasting of 2.5\% VaR and ES performance measured by the standardized score difference.}\label{imp2}
\centering
\tabcolsep=10pt
\begin{tabular}[l]{@{}l | cc|cc}
\toprule
&   \multicolumn{2}{c|}{VaR} &    \multicolumn{2}{c}{ES} \\ \hline
Market & ES-CaViaR & RES-CaViaR& ES-CaViaR & RES-CaViaR\\ \hline
U.S. & 2.04 & 4.49 & 1.67 & 3.91 \\ 
  Germany & 2.06 & 3.16 & -0.82 & 2.71 \\ 
  Hong Kong & 3.93 & 5.03 & 0.89 & 3.20 \\ 
  Japan & 6.37 & 5.66 & 5.80 & 5.60 \\
\bottomrule															
\end{tabular}
\end{table}

\section{Conclusion}
This study offers a combination of the semi-parametric model with realized volatility and the concept
of nowcasting through overnight information for forecasting VaR and ES simultaneously.
We extend a semi-parametric regression model based on asymmetric Laplace distribution and offer a family of RES-CAViaR-oc models by adding overnight return and realized measures as a nowcasting method.
We further employ the adaptive MCMC method in Bayesian inference for parameter estimation and tail forecasting due to the advantage of estimating complex models. We also see optimal convergence for every parameter.
In addition, we conduct comprehensive backtests to ensure forecasting capability of the proposed models in the out-of-sample period.

The empirical study finds that both the RES-CAViaR-oc and RES-CAViaR-oc$^-$ models are more favorable than the original ES-CAViaR,  ES-CAViaR-oc and RES-CAViaR models in terms of the quantile and AL log scores. This suggests that realized volatility and overnight information are two important factors that are useful for predicting tail risk. Murphy diagrams also confirm that CAViaR-type models with realized volatility and overnight returns are more efficient in forecasting tail risk than other models. The results help financial institutions raise capital allocation efficiency, allowing them more profit maximization opportunities.


\section*{Acknowledgement}
We extend our gratitude to the editor, the associate editor, and the anonymous referees for their invaluable time and insightful comments on our paper. Cathy W.S. Chen’s research is funded by the National Science and Technology Council, Taiwan
(NSTC109-2118-M-035-005-MY3 and NSTC112-2118-M-035-001-MY3).
Takaaki Koike is supported by Japan Society for the Promotion of Science (JSPS KAKENHI Grant Number JP21K13275).



\renewcommand{\baselinestretch}{1.5}

\newpage

\appendix

\makeatletter
\renewcommand \thesection{S\@arabic\c@section}
\renewcommand\thetable{S\@arabic\c@table}
\renewcommand \thefigure{S\@arabic\c@figure}
\makeatother

\begin{center}
\bf \Large    Supplement to\\
 ``Tail risk forecasting with semi-parametric regression models by incorporating overnight information''
\end{center}

\begin{figure}[H]
\includegraphics[width=1.0\textwidth]{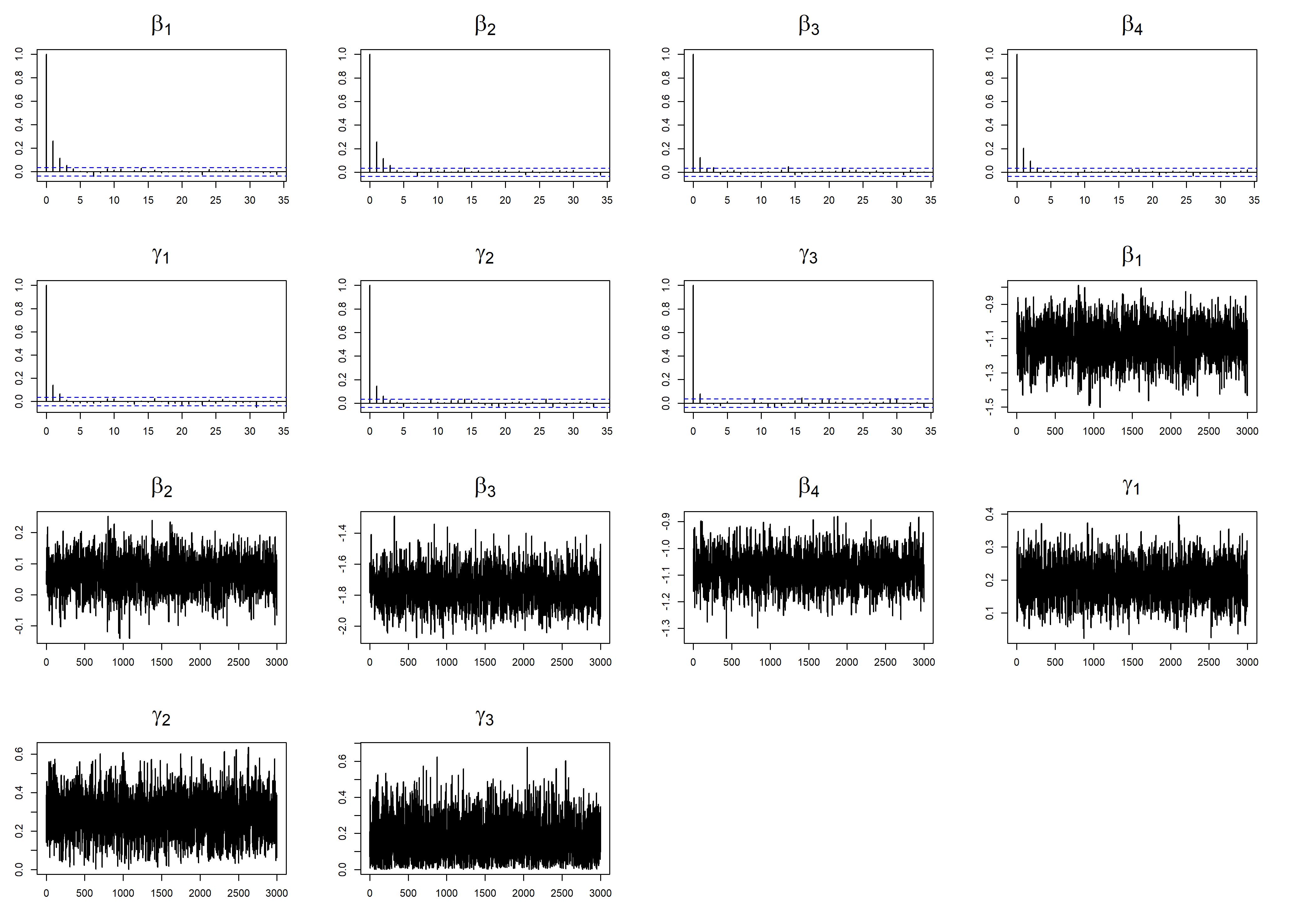}
\caption{ACF and trace plots after the burn-in period for the U.S. market from RES-CAViaR-oc$^-$ model.}
\label{acfus}
\end{figure}

\begin{figure}[H]
\includegraphics[width=1.0\textwidth]{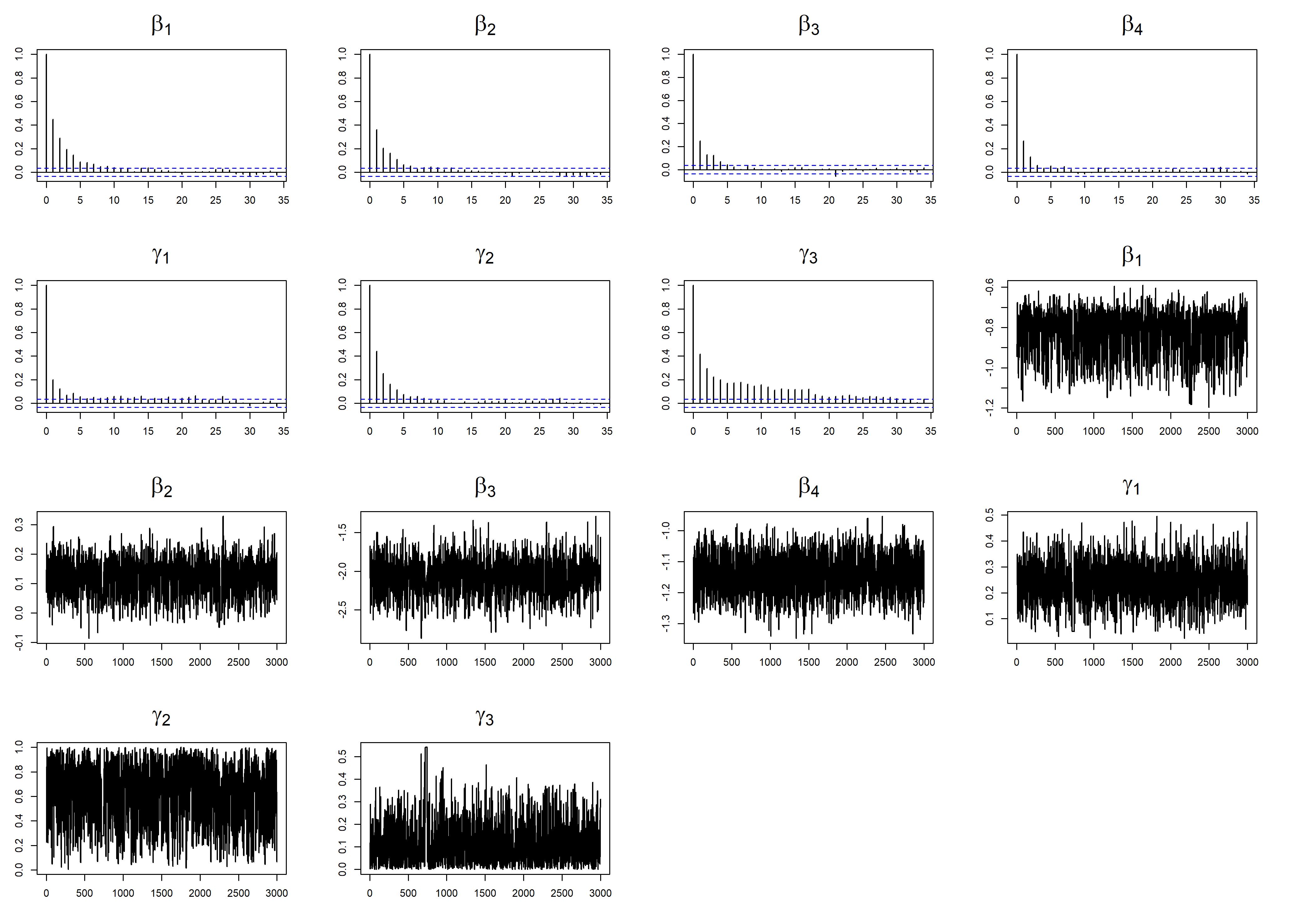}
\caption{ACF and trace plots after the burn-in period for the Germany market from RES-CAViaR-oc$^-$ model.}
\label{acfgermany}
\end{figure}

\begin{figure}[H]
\includegraphics[width=1.0\textwidth]{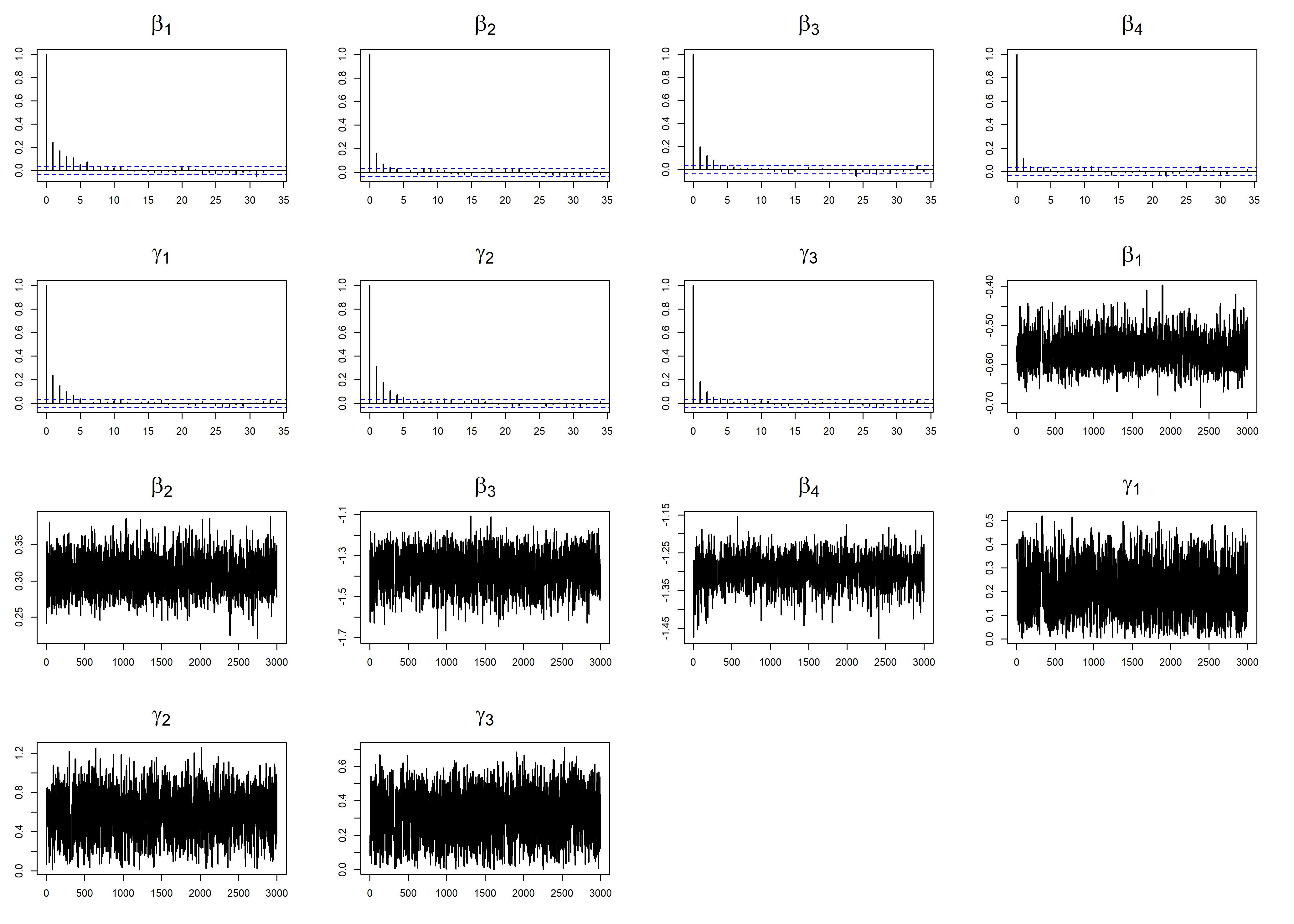}
\caption{ACF and trace plots after the burn-in period for the Hong Kong market from RES-CAViaR-oc$^-$ model.}
\label{acfhk}
\end{figure}

\begin{figure}[H]
\includegraphics[width=1.0\textwidth]{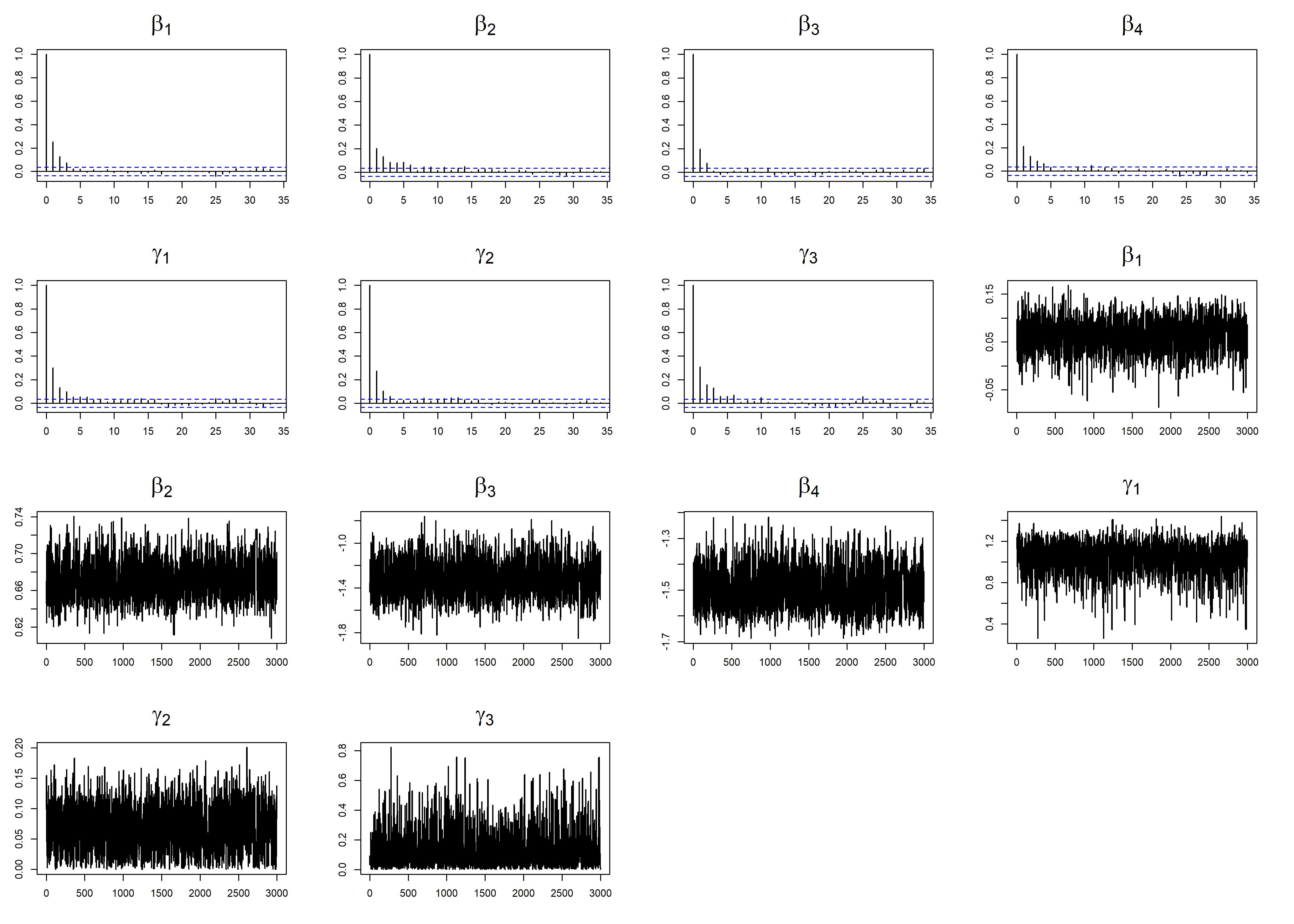}
\caption{ACF and trace plots after the burn-in period for the Japan market from RES-CAViaR-oc$^-$ model.}
\label{acfjp}
\end{figure}

\begin{figure}[H]
\includegraphics[width=1.0\textwidth]{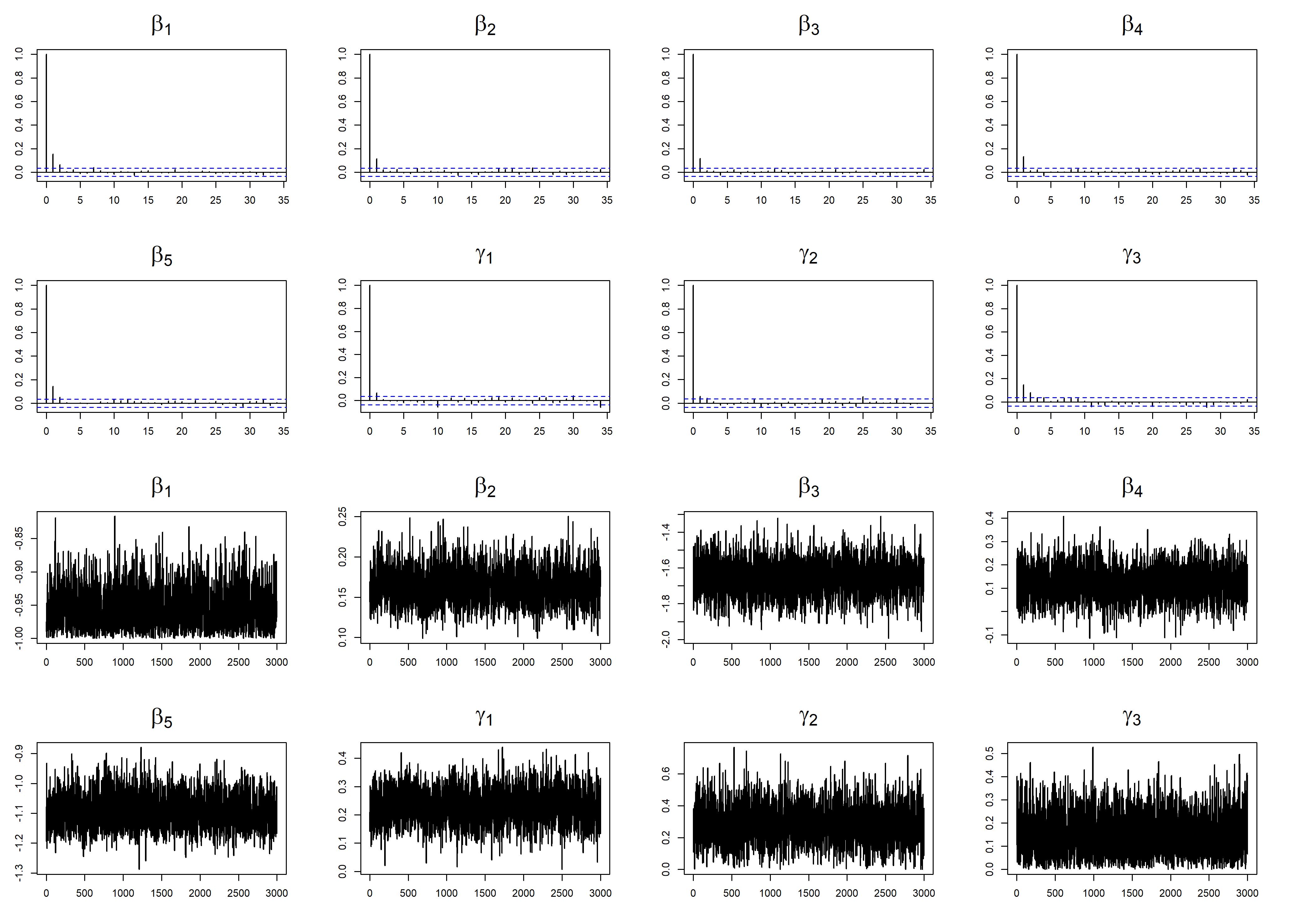}
\caption{ACF and trace plots after the burn-in period for the U.S. market from RES-CAViaR-oc model.}
\label{acf1us}
\end{figure}

\begin{figure}[H]
\includegraphics[width=1.0\textwidth]{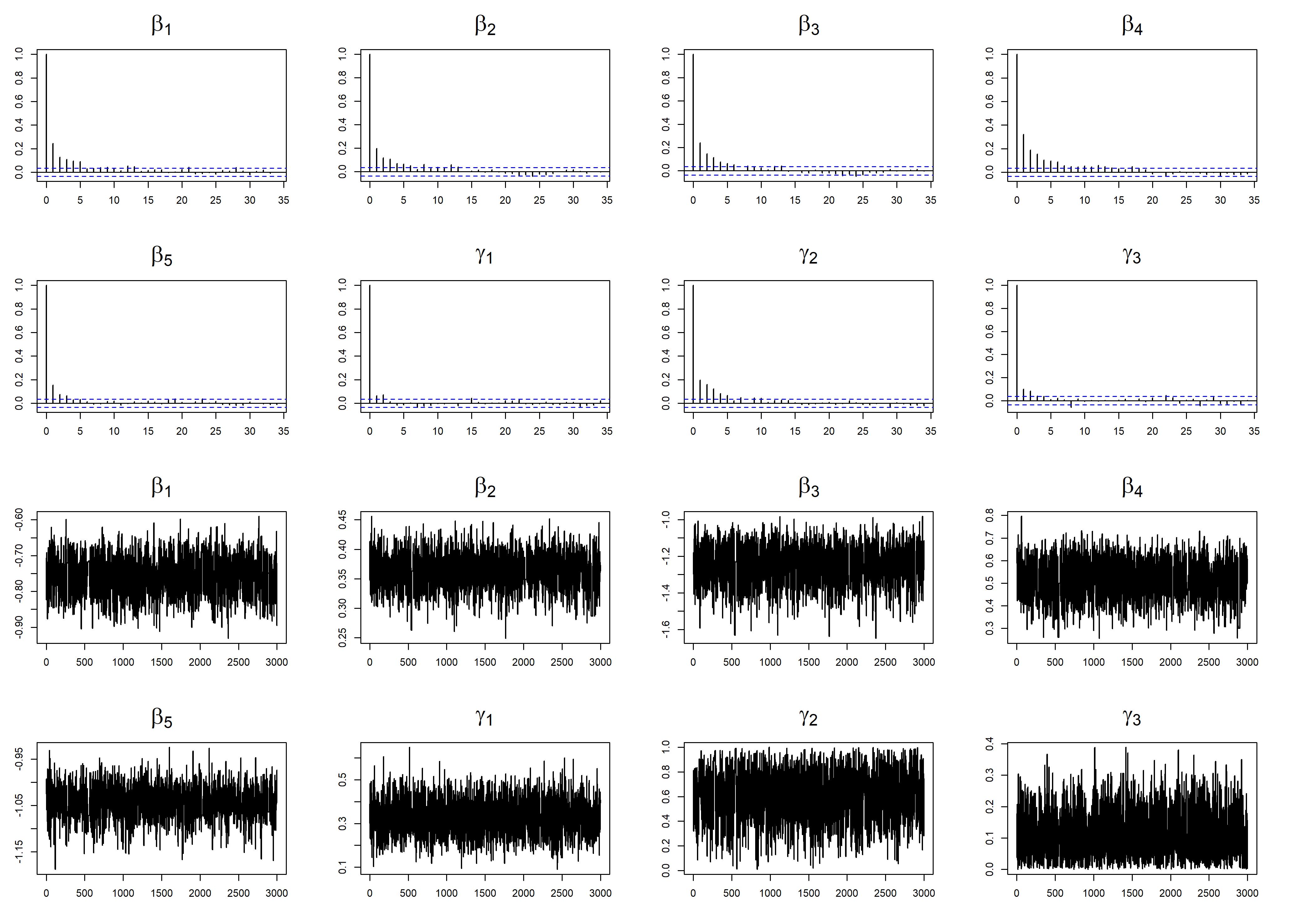}
\caption{ACF and trace plots after the burn-in period for the Germany market from RES-CAViaR-oc model.}
\label{acf1germany}
\end{figure}

\begin{figure}[H]
\includegraphics[width=1.0\textwidth]{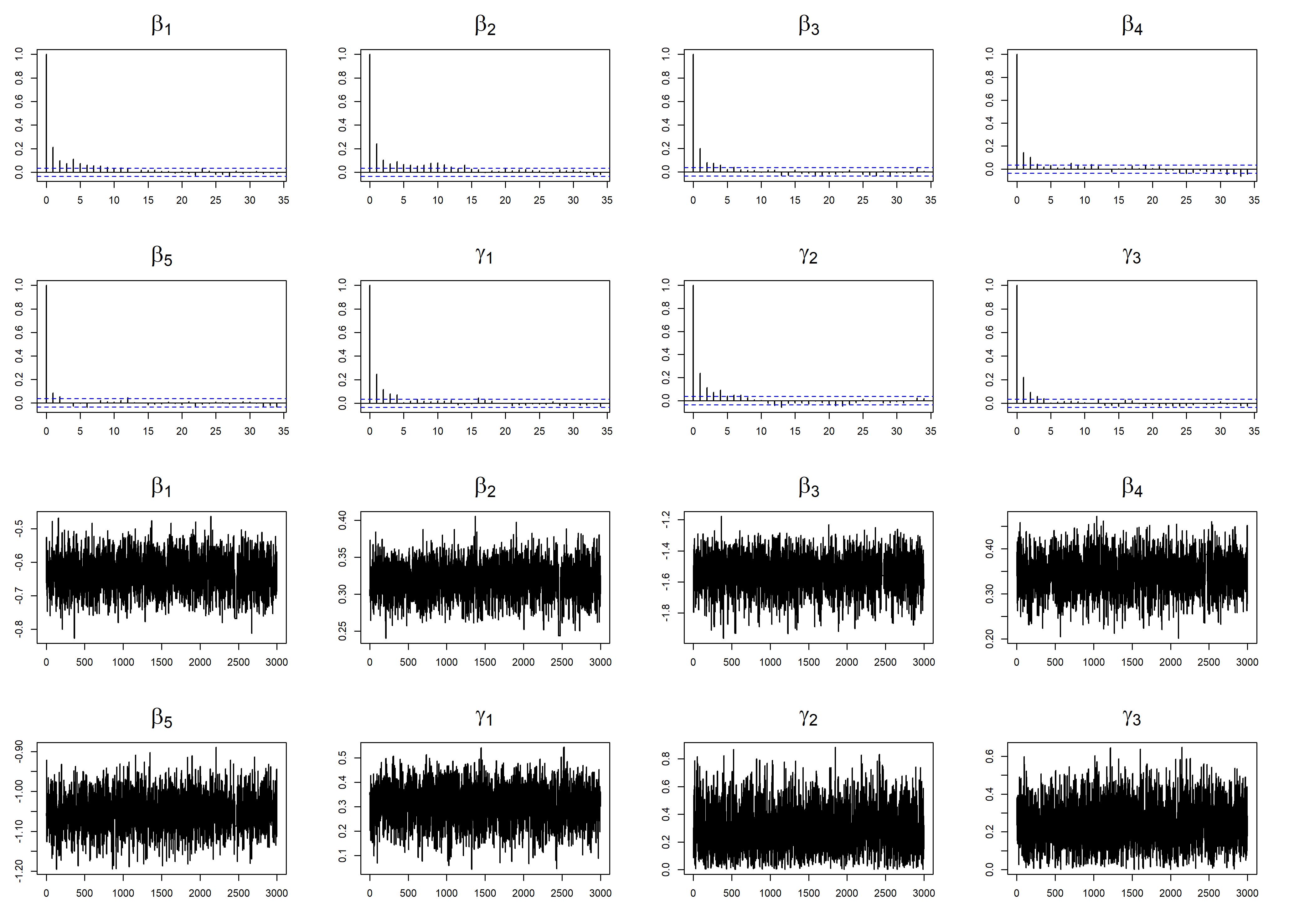}
\caption{ACF and trace plots after the burn-in period for the Hong Kong market from RES-CAViaR-oc model.}
\label{acf1hk}
\end{figure}

\begin{figure}[H]
\includegraphics[width=1.0\textwidth]{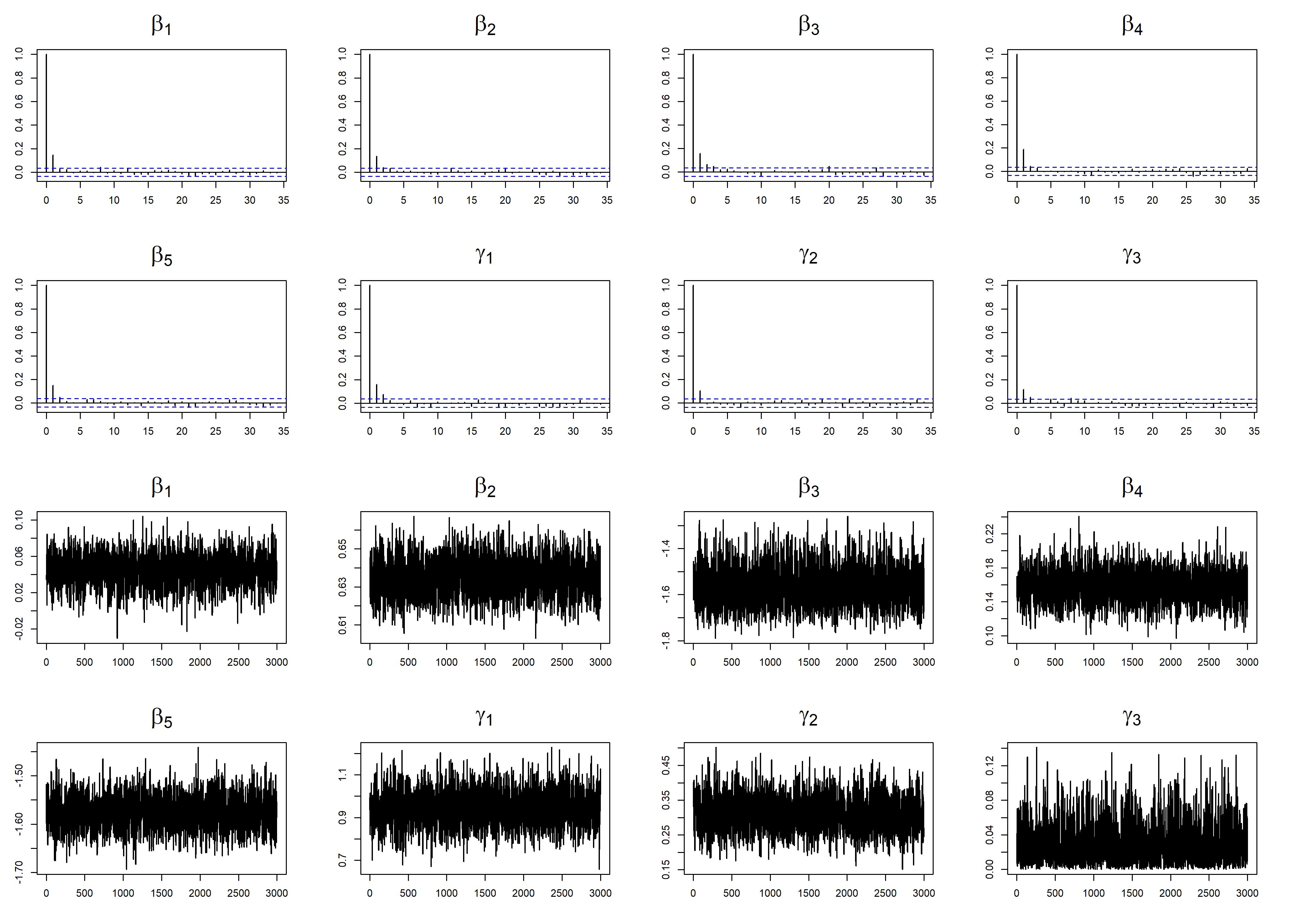}
\caption{ACF and trace plots after the burn-in period for the Japan market from RES-CAViaR-oc model.}
\label{acf1jp}
\end{figure}

To monitor the convergence and stability of MCMC iterates for the stock markets of the U.S., Germany, Hong Kong, and Japan, Figures \ref{acfus} to \ref{acfjp} present ACF plots and trace plots for each parameter based on the RES-CAViaR-oc$^{-}$ model, while Figures \ref{acf1us} to \ref{acf1jp} focus on the RES-CAViaR-oc model. We carry out $M=20,000$ MCMC iterations, discard the first $N=8,000$ as the burn-in period, and include only every fourth iteration in the sample period for inference.

\begin{figure}[H]
\centering
\scalebox{0.8}{
\includegraphics[width=0.8\textwidth]{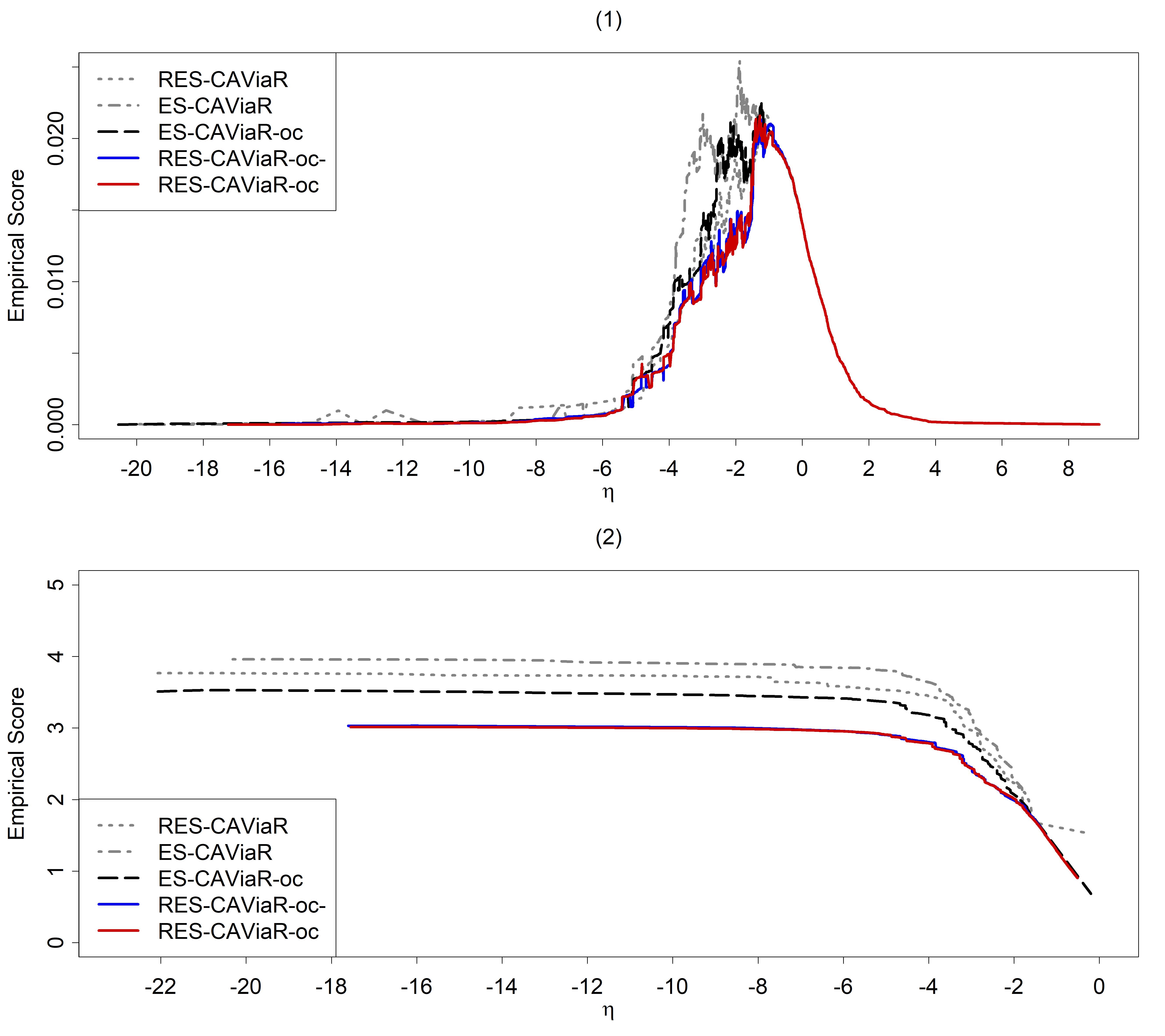}
}\caption{Murphy diagrams for (1) VaR and (2) ES at the $2.5\%$ level for the U.S. market.}
\label{Fig9}
\end{figure}

\begin{figure}[H]
\centering
\scalebox{0.8}{
\includegraphics[width=0.8\textwidth]{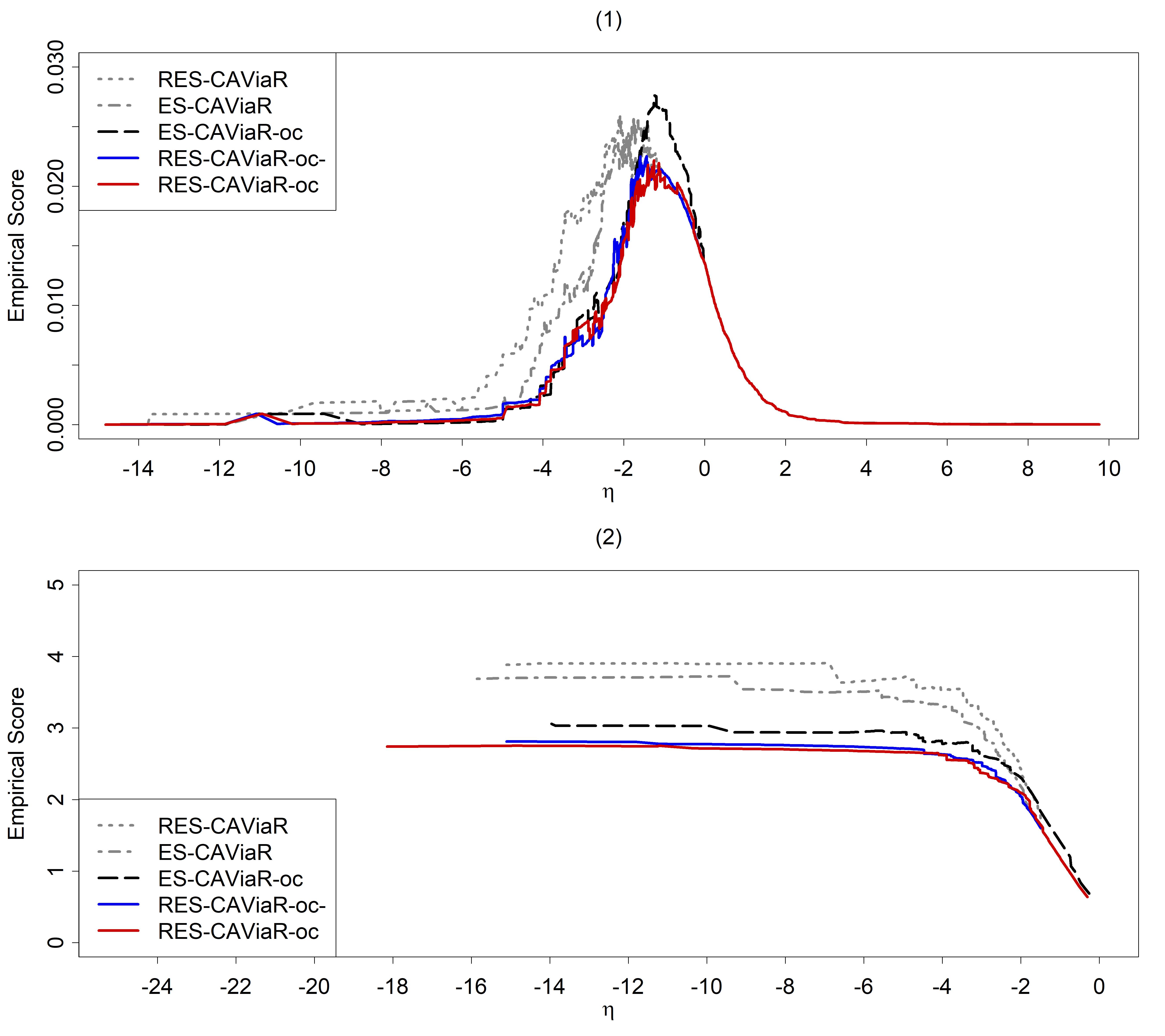}
}\caption{Murphy diagrams for (1) VaR and (2) ES at the $2.5\%$ level for the Germany market.}
\label{Fig10}
\end{figure}

\begin{figure}[H]
\centering
\scalebox{0.8}{
\includegraphics[width=0.8\textwidth]{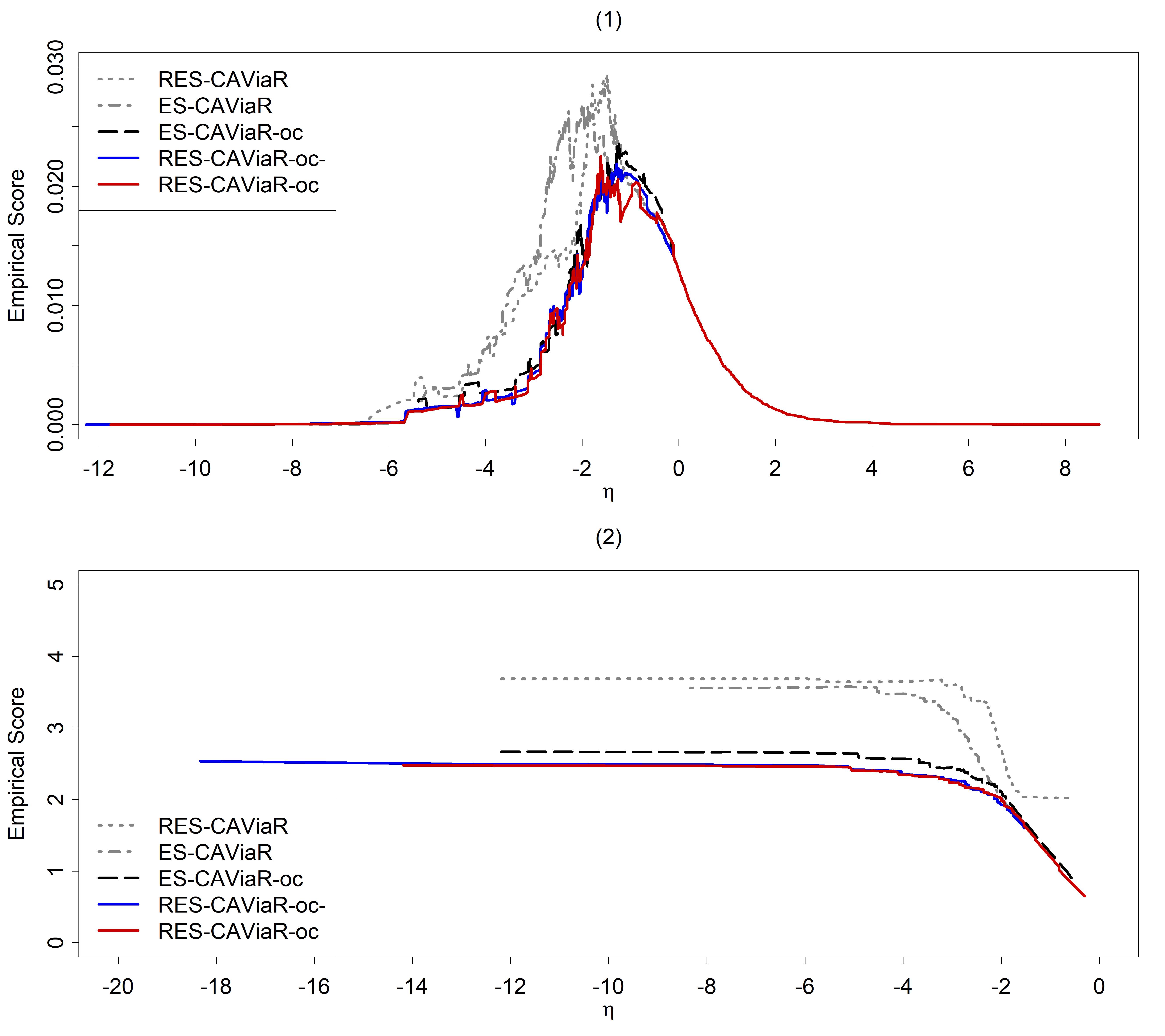}
}\caption{Murphy diagrams for (1) VaR and (2) ES at the $2.5\%$ level for the Hong Kong market.}
\label{Fig11}
\end{figure}

\begin{figure}[H]
\centering
\scalebox{0.8}{
\includegraphics[width=0.8\textwidth]{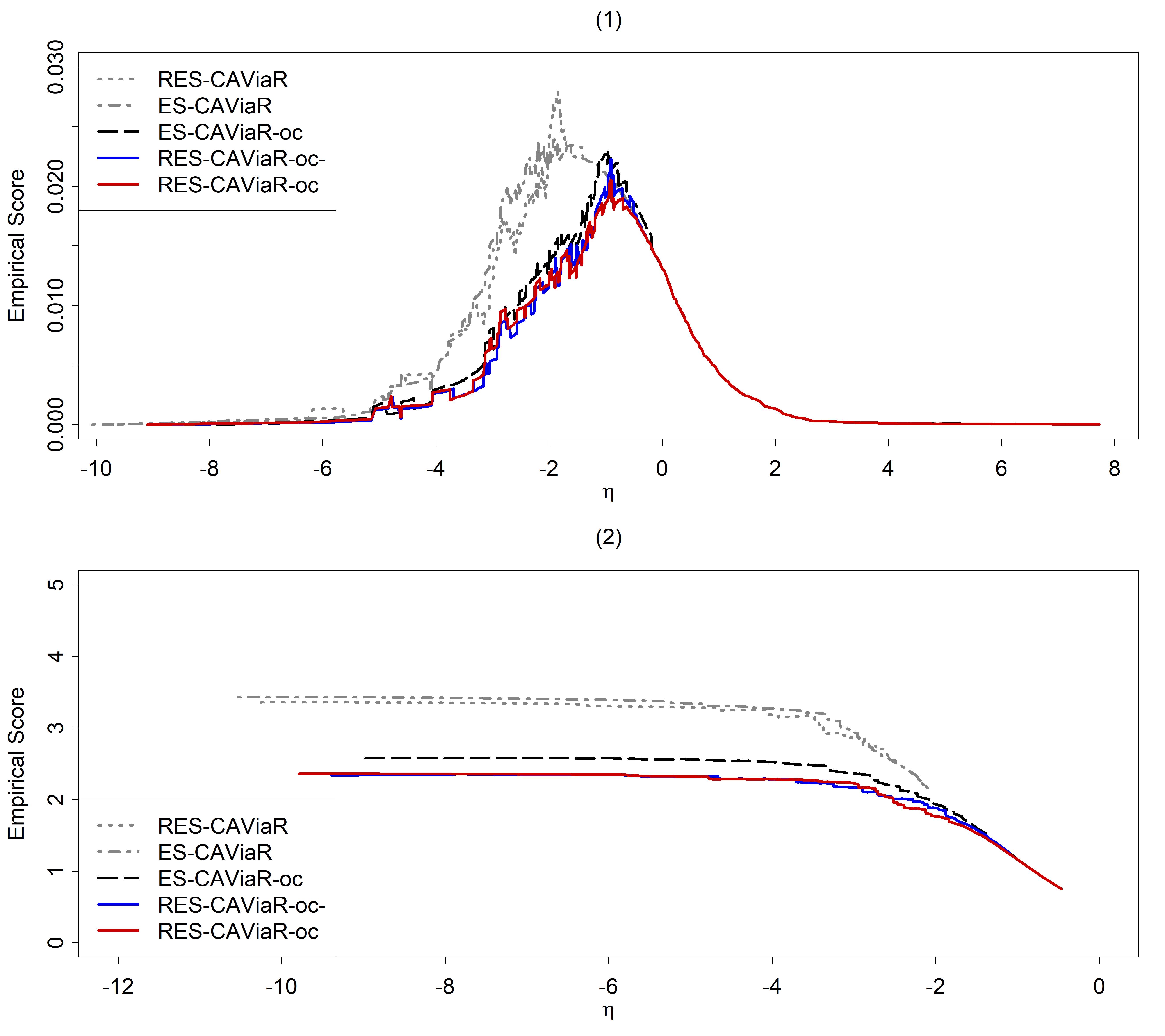}
}
\caption{Murphy diagrams for (1) VaR and (2) ES at the $2.5\%$ level for the Japan market.}
\label{Fig12}
\end{figure}

Figures \ref{Fig9} through \ref{Fig12} display Murphy diagrams for (1) VaR and (2) ES at the 2.5\% level for the stock markets of the U.S., Germany, Hong Kong, and Japan.


\end{document}